\documentstyle[emulateapj,psfig]{article}
\font\gkvec=cmmib10                         %for boldface lowercase Greek
\def\bzeta{\hbox{{\gkvec\char16}}}        %bold face zeta
\lefthead{Agol \& Krolik}
\righthead{Quasar Microlensing}
\begin{document}

\title{Imaging a Quasar Accretion Disk with Microlensing}

\author{Eric Agol and Julian Krolik}
\affil{Physics and Astronomy Department, Johns Hopkins University,
    Baltimore, MD 21218}
\begin{abstract}
      We show how analysis of a quasar high-magnification microlensing event 
may be used to
construct a map of the frequency-dependent surface brightness of the
quasar accretion disk.  The same procedure also allows determination of the
disk inclination angle, the black hole mass (modulo the caustic velocity),
and possibly the black hole spin.  This method depends on the validity of
one assumption: that the optical and ultraviolet continuum of the
quasar is produced on the surface of an azimuthally symmetric, flat 
equatorial disk, whose gas follows prograde circular orbits in a Kerr 
spacetime (and plunges inside
the marginally stable orbit).  Given this assumption, we advocate using
a variant of first-order linear regularization to invert multi-frequency
microlensing lightcurves to obtain the disk surface brightness as a
function of radius and frequency.  The other parameters can be found by
minimizing $\chi^2$ in a fashion consistent with the regularized solution
for the surface brightness.

      We present simulations for a disk model appropriate to the Einstein
Cross quasar, an object uniquely well-suited to this approach.   These
simulations confirm that the surface brightness can be reconstructed
quite well near its peak, and that there are no systematic errors in
determining the other model parameters.  We also discuss the observational
requirements for successful implementation of this technique.
\end{abstract}


\section{Introduction}

Due to their great distance and small intrinsic size, it is not possible to
obtain a resolved optical image of a quasar with current technology.  The
angular size of a quasar optical emission region is of
order ${10^{-8}}^{\prime\prime}$, a scale so small as to require a baseline of 
several thousand kilometers to resolve.  Thus, until optical
VLBI becomes practical, quasar structure will need to be probed by
other, indirect, means.  Reverberation mapping provides an instructive
example of the difficulties of such indirect approaches, for it
has proven difficult both to implement and to interpret.  We believe that
the subject of this paper, microlensing by stars in an intervening
galaxy, is a more promising method.

The circumstantial evidence that black holes power quasars is convincing: 
accretion onto black holes can be very efficient in converting rest mass
energy to photons (up to 40\%) or to bulk momentum (forming radio jets/lobes)
in a compact region which has an effective temperature near where the quasar
spectrum peaks.  There is statistical evidence that quiescent black holes
in the nuclei of galaxies at low redshift could be the remnants of quasars.
And, quasars have properties very similar to Seyfert galaxies, for some of which
there is good spectroscopic evidence for a central black hole.  The nature
of the accretion flow is quite uncertain, however, although it is probably
geometrically thin since angular momentum will support the accretion flow
against collapse and geometrically thick disks are by nature inefficient
(if quasar disks were thick, there would be problems producing the huge
luminosities observed within a reasonable mass budget).

   Attempts to constrain the character of the innermost accretion flow
by means of spectral modeling have made little progress for several
reasons.  There are major systematic uncertainties about fundamental issues
(e.g., the vertical distribution of energy dissipation, the physics
to include in radiation transfer solutions).  In addition, any
particular model depends on a sizable number of free parameters (mass
of the central black hole, accretion rate, viscosity parameter,
inclination angle), so that parameter estimation is tricky.

     Analysis of a microlensing event is potentially a more powerful tool.
Rather than guess a specific model for the accretion disk, the history
of magnification in the event can be used to directly infer the disk surface
brightness as a function of radius and frequency.  The only assumptions
required are that the continuum emission surface is geometrically flat,
the material forming it follows circular orbits, and general relativity
determines the dynamics of both the matter and the photons.

     To apply this technique requires study of a particular
gravitationally-lensed quasar.  Many are now known, and some appear
to undergo occasional fluctuations due to individual stars in the
lens galaxy magnifying the quasar, a phenomenon referred to as microlensing.
The Einstein Cross is particularly well-suited to this problem for
a number of reasons that we will detail in \S 1.3.  Several authors have
attempted to constrain the character of the accretion disk in this
system by comparing predicted lightcurves to the data compiled during
microlensing events (Jaroszy\'nski et al. 1992, Rauch \& Blandford 1991,
Jaroszy\'nski \& Marck 1994, Czerny et al. 1994).  However, the results have
all been somewhat inconclusive due to the ordinary spectral modeling
difficulties described above.

    The remainder of this paper is organized as follows: In \S 1, we
describe how a caustic crossing can be used to perform the mapping,
and discuss why the Einstein Cross is such a suitable target for
this sort of study.  In \S 2, we discuss the model in detail, considering our
assumptions, the inversion technique, and error propagation.
In \S 3, we present a variety of simulations of caustic-crossing events, 
and demonstrate that we can measure most model parameters.  We show some
examples of the accuracy with which the intensity at the accretion disk may
be determined, and discuss how the reliability of the results depends
on the quality of the observational data, on the character of the
monitoring, and, to some degree, on the character of the regularization
scheme.  In the final section, we discuss the results
and present our conclusions.

\subsection{Smooth thin disk}

For a standard thin accretion disk with constant accretion rate and no advection
of heat, the energy generation per unit area as a function of radius is given by
\begin{equation}\label{energy}
Q = {3 \over 4\pi}{ GM_{BH} \dot M \over r^3} R_R (r).
\end{equation}
where $M_{BH}$ is the mass of the black hole, $\dot M$ is the accretion
rate, $r$ is the radius within the disk, and $R_R$ is a
correction factor that combines outward advection of energy associated with the
angular momentum flux and relativistic effects (Page \& Thorne 1974,
notation from Krolik 1998).  $R_R$ is a
function of the black hole spin $a_s = a/M_{BH}$.

Though equation~[\ref{energy}] describes the functional dependence of the 
energy released with radius, it does not specify whether this energy is thermal
or mechanical, and does not specify the dissipation as a function of height
within the accretion disk.  The appearance of a standard thin accretion 
disk can vary significantly depending on how and where the energy is released.
To further our understanding, it would be very desirable to actually measure
the local spectrum.

   One potential method to achieve this is to observe a quasar 
during the sort of high-magnification microlensing event that occurs when
a caustic crosses the source (Grieger et al., 1988 and 1991, Gould \& Gaudi
1997).  Grieger et al. (1991) advocate inverting the microlensing lightcurve
to obtain the one-dimensional surface brightness of the quasar,
$P_\nu(x)=\int I_\nu(x,y) dy$, where $I_\nu(x,y)$ is the specific intensity
of the quasar at sky coordinates $(x,y)$.
Their method is quite elegant, but relies on first-order regularization
(the assumption that the quasar profile is smooth).  This assumption is
problematic for black hole models since the energy release increases
rapidly towards smaller radii, and relativistic effects can cause sharp peaks in
the quasar profile.  These sharp features are smoothed out when this method
is used (see Figure 1).

   We believe a better approach is to solve instead for the surface brightness 
at the accretion disk, with the relatively benign assumptions that the disk 
is planar, axisymmetric, and isotropically emitting (in the fluid frame), and 
most importantly, likely to vary smoothly with radius.  This approach
provides a better inversion of the quasar profile if the disk assumptions
are correct. This method recovers the spectrum as a function of radius at
the accretion disk, which can then be compared with disk atmosphere models or 
other spectral modeling.  In addition, it can help constrain disk parameters, 
such as the inclination angle, without relying on a specific accretion disk 
spectral model.

\vskip 2mm
\hbox{~}
\centerline{\psfig{file=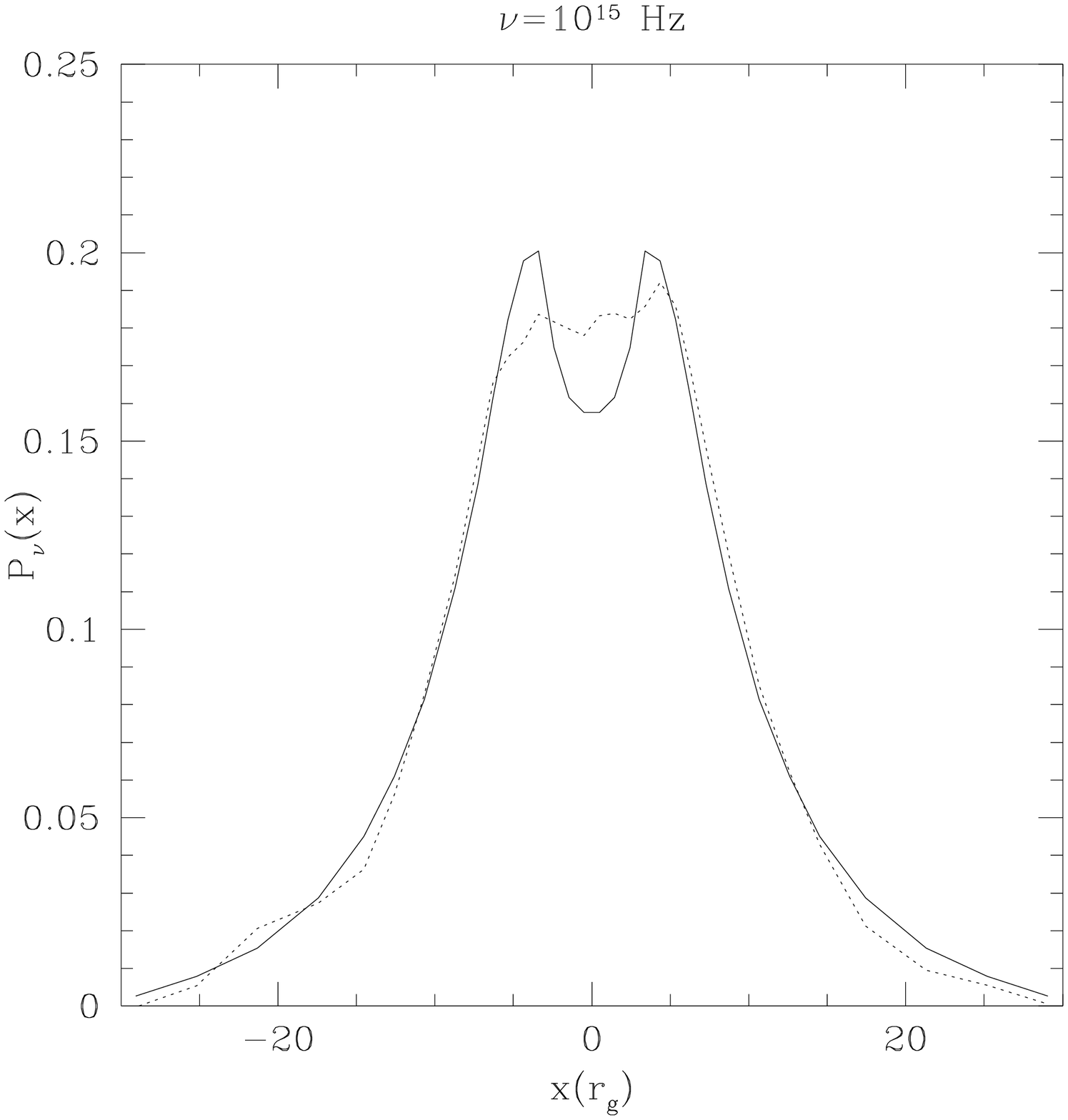,width=3.7in}} %FIGURE 1
\noindent{
\scriptsize \addtolength{\baselineskip}{-3pt}
\vskip 1mm
\begin{normalsize}
Fig.~1.\
%\caption{
Plot of recovered one-dimensional disk profile as a function of impact
parameter for a face-on accretion disk using the Grieger et al. (1991) 
technique.  Solid line is the model profile; dotted line is recovered 
profile.  The parameter $\alpha$ is distance across the source 
perpendicular to the caustic line in units
of gravitational radii. %\label{fig01}
\end{normalsize}
\vskip 3mm
\addtolength{\baselineskip}{3pt}
}

\subsection{High magnification microlensing event}

The spectrum of a quasar as a function of time is a convolution of the lensing 
magnification with the surface brightness of the quasar. Near a fold caustic, 
two bright images merge (Schneider et  al. 1992).  The resulting magnification 
has the specific form:
\begin{equation}\label{camp}
A(x,t)=A_0 + {K \over \sqrt{x-v_c(t-t_0)}} \Theta\left[x-v_c(t-t_0)\right],
\end{equation}
where $A_0$ is the magnification due to the additional images, $x$ is the
position on the quasar plane perpendicular to the caustic, $K$ is the
strength of the caustic [units of $({\rm distance})^{1/2}$], $v_c$ is the
speed of the caustic (assumed to be positive),
$\Theta$ is the step function [$\Theta(x)=1$ for $x>0$, $0$ otherwise], 
and $t_0$ is the time at which the caustic crosses the $x=0$ point measured 
relative to the center of the lightcurve (Schneider \& Wei\ss\ 1986).
This equation is valid whenever the source size is small compared
to the Einstein ring radius of the microlensing star.  If that ratio is 
small, there is only a small probability of any of a
number of problems that might invalidate equation~[\ref{camp}].  The list
of potential problems includes: 1) the possibility that the source is
projected behind a cusp (where the caustic curve discontinuously changes
direction) or behind the crossing of multiple fold caustics;
2) significant curvature in the caustic; 3) variation of $K$ 
along the caustic on the scale of the source.  Grieger et al. (1988) 
demonstrate that a source of size smaller than $\sim 10$\% of the Einstein 
radius of the typical lensing mass is necessary for the second and 
third assumptions to be valid.  

\subsection{Microlensing Laboratory:  The Einstein Cross}

The Einstein Cross is a quasar (z=1.695) lensed into four images
by a nearby (z=0.0394) barred spiral galaxy (Huchra et al. 1985).
The luminosity distance to the quasar is $D_L=3\times 10^{28} 
h_{75}^{-1}$ cm, where $h_{75}=H_o$/(75 km/s/Mpc), assuming
$\Omega_m=1,\Lambda=0$.

   This quasar has been observed to undergo fluctuations due to
microlensing roughly once per year (Irwin et al. 1989; R. Webster, private
communication).  It is particularly well suited for studying
microlensing because the lensing galaxy is nearby.  This happy
coincidence makes the time delays between the four images all
less than a day, so that it is easy to distinguish
intrinsic variability from microlensing variability.  It also
makes the stellar velocities projected onto the source plane quite
high, so that microlensing is frequent, and also makes individual events
comparatively brief (only about a month, in contrast to the
years to decade timescales characteristic of more distant lenses).
In addition, the Einstein radius projected onto
the quasar plane is quite large compared to
the quasar size, validating the assumptions made in the previous
subsection, and also making the microlensing variations strong.

The models of Witt et al. (1993) suggest that for image A,
\begin{equation}
\langle K \rangle \simeq 8 \left({r_E\over 5.8\times 10^{16}{\rm cm}}
\right)^{1\over 2} M_9^{-{1\over 2}}, 
\end{equation}
where $r_E=5.8\times 10^{16}{\rm cm}(m/0.2)^{1/2} h_{75}^{-1/2}$
is the Einstein radius projected to the quasar plane (Schneider et
al. 1992), $m$ is the typical mass (in Solar units) of a star causing
microlensing, and
$M_9=M_{BH}/10^9 M_\odot$.  In equation (3), and in the rest of the paper, 
we adopt $r_g \equiv GM/c^2$ as the unit of distance.  For concreteness,
we will use $K=8$ in our simulations.  The parameter $A_0$ can be
approximated by $\langle A_0 \rangle = |(1-\sigma)^2-\gamma^2|^{-1}$
where $\gamma$ is the shear (Witt et al. 1993).  
%referee change 2
For image A, the
estimated range of the microlensing parameters is $\sigma=0.3-0.4$ and
$\gamma=0.4-0.5$, giving $A_0=3-9$; we use $A_0=6$ in our simulations.
%end change 2

\section{Microlensing of an Accretion Disk}

\subsection{Caustic crossing}

As the caustic crosses the quasar, the observed lightcurve is a
convolution of the magnification, equation [\ref{camp}] and 
$P_\nu(x)$:
\begin{equation}\label{eqonedim}
F_\nu(t)=\int_{-\infty}^{\infty} A(x,t) P_\nu(x) dx.
\end{equation}
Note that if the flux within a waveband
can be measured outside of the caustic and subtracted off, then
the dependence on $A_0$ disappears, and the $K$ parameter becomes degenerate
with an arbitrary scaling of $P_\nu(x)$.
As discussed by Grieger et al. (1991), equation~[\ref{eqonedim}] can be
inverted using regularization to find $P_\nu(x)$.  
Similar techniques have been used for measuring the limb-darkening
of stars during galactic microlensing events 
(Gaudi \& Gould 1998, Albrow et al. 1998).

\subsection{Black Hole geometry}

Near a black hole, relativistic effects cause Doppler beaming of the
emitted radiation, gravitational red shifts, and bending of photon trajectories.  
To image the surface of an accretion disk, these relativistic effects must be
accounted for using a relativistic transfer function (defined in Cunningham 
1975).  To compute the transfer function (see equation [\ref{fint}] below), 
we make several simplifying assumptions: (1) the accretion disk is 
thin, i.e. $h \ll r$; (2) the gas follows prograde circular orbits outside
the marginally stable radius $r_{ms}$, and undergoes freefall within 
$r_{ms}$ with constant angular momentum and energy equal to those
obtaining at $r_{ms}$; (3) the disk is flat and
lies in the equatorial plane of the black hole; (4) the gas emits 
isotropically in its rest frame, i.e.,
there is no limb darkening in the accretion disk atmosphere.   The
first two assumptions are appropriate if pressure gradients cause forces
much smaller than the gravitational force in the $z$ and $r$ directions,
respectively.  This condition is not met in advection-dominated accretion
flows or slim accretion disks (Beloborodov 1998).  In the case of slim 
accretion disks,
the orbital frequency is nearly Keplerian, and deviates by less than
20\% when $\dot M c^2/L_{Edd}\leq 1000$; however, the disk scale height
can become a large fraction of the radius, which changes the emitted
angle of radiation relative to the disk normal and can cause shadowing 
which we do not take into account.  The third assumption is inappropriate
if the disk is warped; however, Bardeen-Petterson precession (1975)
can align the disk and black hole by the time the gas reaches the inner
radii.  The fourth assumption is a simplification for greater ease in
the inversion computation since the disk can be viewed from only one angle
and thus at most one emitted angle can be observed at each 
radius/azimuth of the disk.  For each radius, there is a limited range
of emitted angles which are observed, so our inversion will give
some sort of average of the intensities within that range.
\vskip 2mm
\hbox{~}
\centerline{\psfig{file=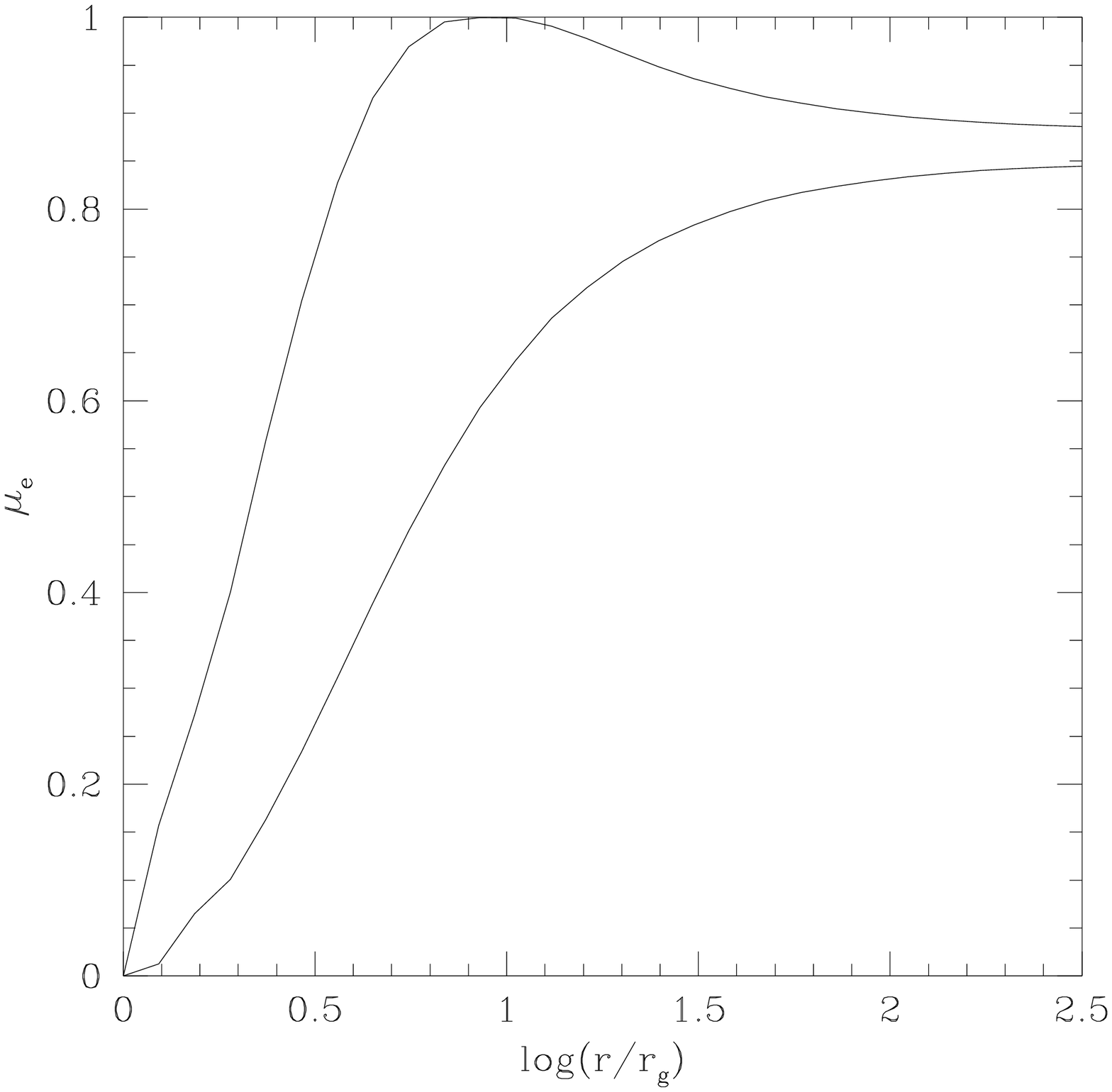,width=3.7in}}
\noindent{
\scriptsize \addtolength{\baselineskip}{-3pt}
\vskip 1mm
\begin{normalsize}
Fig.~2.\ The upper curve shows the maximum $\mu_e=\cos{\theta}$, where 
$\theta$ is the angle between the disk normal and the direction of the 
photon in the fluid rest frame. The lower curve shows the minimum 
$\mu_e$. The disk parameters are $i=30^\circ$ and $a_s=0.998$. %\label{fig02}
\end{normalsize}
\vskip 3mm
\addtolength{\baselineskip}{3pt}
}
Figure 2 shows the range of emitted angles (in the fluid
rest frame, $\mu_e$ is the cosine of the normal to the disk) for a disk 
inclined at
30$^\circ$.   For a face-on disk, only one emitted angle is seen at 
each radius for all azimuths, so this assumption simply corresponds to 
mapping the specific intensity
of the disk at $\mu_e(r)$.  We could have assumed some limb-darkening
law, but this is not warranted by the crudeness of the inversion technique.

Figure 3 shows the disk geometry. 
The inclination angle of the accretion disk, $\mu=\cos{i}$, is $i=0^\circ$ when the disk is
face-on, and 90$^\circ$ when the disk is edge-on.  The caustic crossing angle $\phi_c$ is
measured with respect to the $(\alpha,\beta)$ coordinates, which are defined so
that the $\beta$ coordinate lies parallel to the projection of the disk axis onto the 
sky plane (in Figure 3, the disk spin axis is pointing out of the page), with the 
black hole at the origin.  We will use units of $r_g$ for the $(\alpha,\beta)$ coordinates.
\vskip 2mm
\hbox{~}
\centerline{\psfig{file=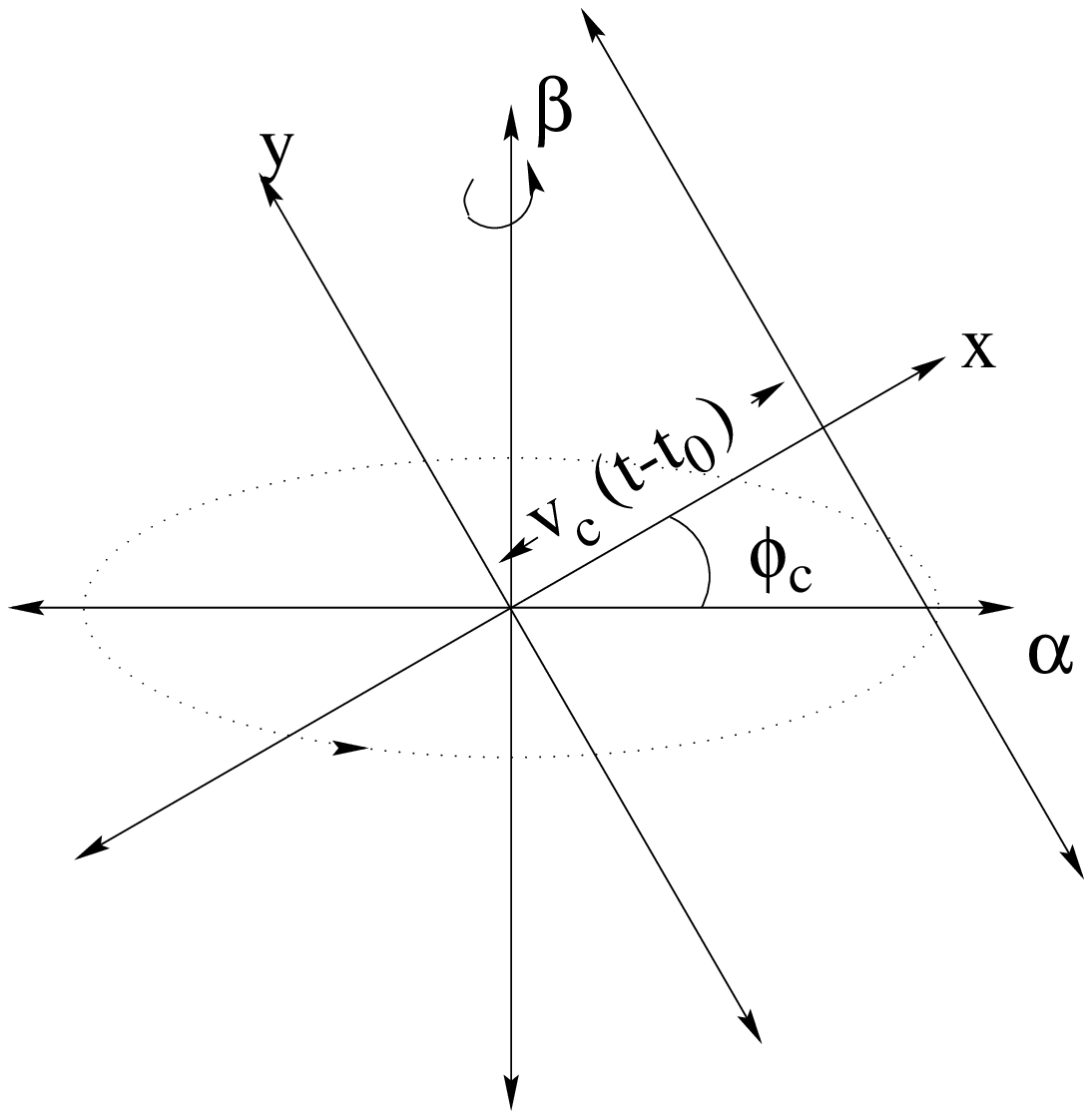,width=4in}}
\noindent{
\scriptsize \addtolength{\baselineskip}{-3pt}
\vskip 1mm
\begin{normalsize}
Fig.~3.\ Geometry of the accretion disk.  The disk axis points up out
of the page, while the $\beta$ axis lies in the page.  The line parallel
to the $y$ axis is the caustic. %\label{fig03}
\end{normalsize}
\vskip 3mm
\addtolength{\baselineskip}{3pt}
}

The rotation of the accretion disk causes beaming of the radiation, leading to
a hot spot on the approaching side, and a cold spot on the receding side.  For 
an exactly face-on disk, the disk is symmetric, so no hot/cold spots exist, but
a dip occurs inside the inner edge of the disk.  
In Figure 4 we show $P_\nu(x)$ for
a blackbody accretion disk at two frequencies.  As the disk
becomes more edge-on, the profile becomes more asymmetric as the Doppler aberration 
becomes stronger.
If the temperature of the disk decreases outwards, then the size of the hot
spot will increase for smaller frequencies, and become less asymmetric, as can
be seen by comparing Figure 4(a) and 4(b).  Figures 4(c)
and 4(d) show $P_\nu(x)$ for different caustic crossing angles,
showing the hotspot is oblong.
\vskip 2mm
\hbox{~}
\centerline{\psfig{file=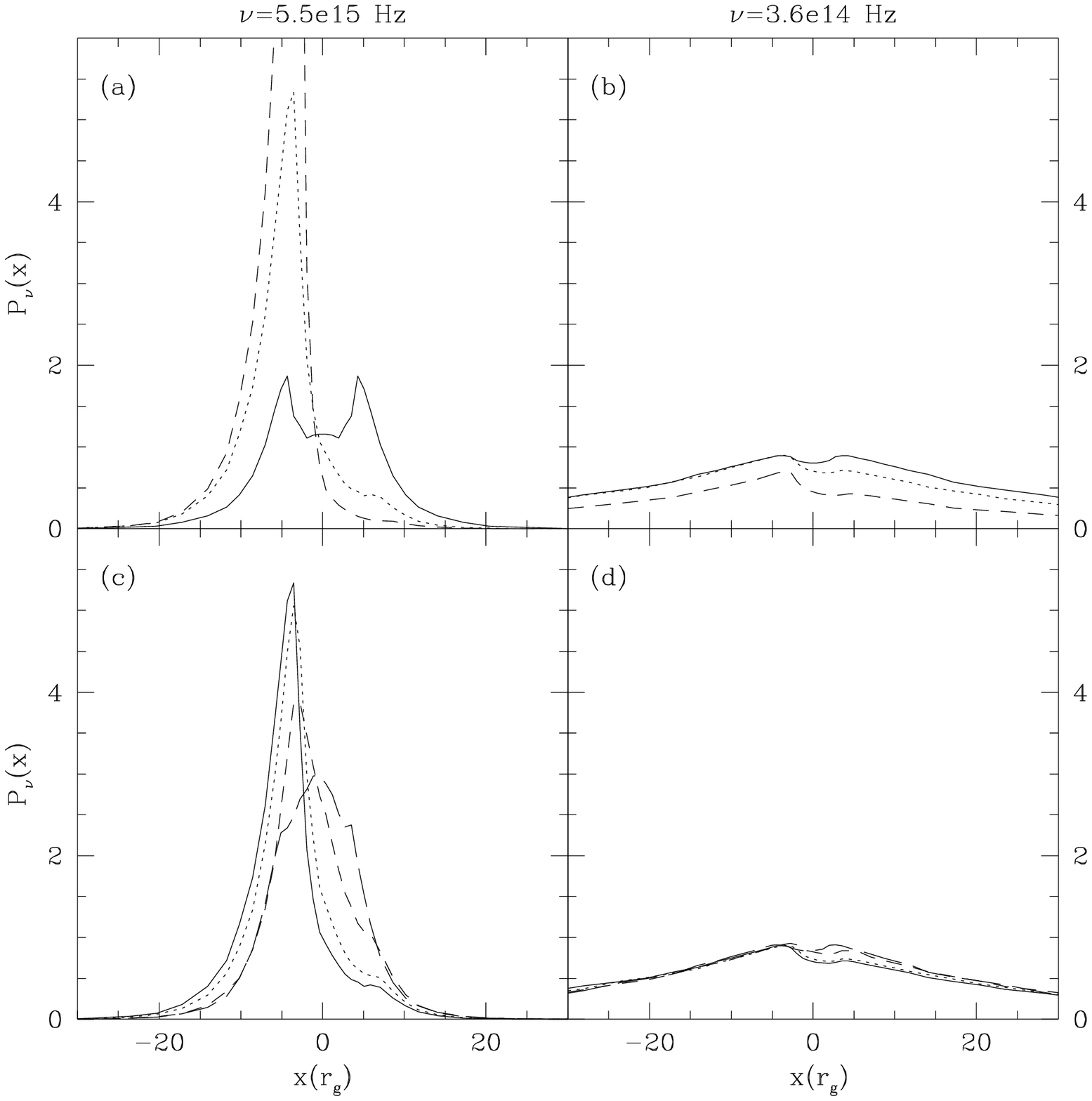,width=3.7in}}
\noindent{
\scriptsize \addtolength{\baselineskip}{-3pt}
\vskip 1mm
\begin{normalsize}
Fig.~1.\ Plots of the disk 1-D profile for a disk with $M_9=1$, $\dot m = 1$, 
and $a_s=0.998$.  Figures (a) and (b) compare inclination angle: 
$i=$ 0$^\circ$ (solid), 30$^\circ$ (dotted) and 60$^\circ$ (dashed) for 
$\phi_c=0$. Figures (c) and (d) compare $\phi_c=$ 0 (solid), 30$^\circ$ 
(dotted), 60$^\circ$ (short dashed), 90$^\circ$ (long dashed) for 
$i=30^\circ$.  The frequencies are in the quasar rest frame. %\label{fig04}
\end{normalsize}
\vskip 3mm
\addtolength{\baselineskip}{3pt}
}

\subsection{Prediction of the Lightcurve}

With these assumptions and parameter choices, the observed flux may be
predicted from the run of intensity with radius:
\begin{equation}\label{fint}
F(\nu_o,t)=\int d\alpha d\beta A\left[x(\alpha,\beta),t\right] g^3 I(\nu_o/g,r_e),
\end{equation}
where $g = \nu_o/\nu_e$ is the redshift between the observer and the emitter
and $I(\nu_e,r)$ is the specific intensity at the accretion disk (we have
assumed it is independent of the emitted angle).  The relation between the 
caustic coordinate and the black hole coordinates is $x(\alpha,\beta)
=\alpha\cos{\phi_c}+\beta\sin{\phi_c}$ (see Figure 3).  The 
physical variables $\nu_e$ and
$r_e$ may be related via a Jacobian to $\alpha$ and $\beta$ through their
functional dependence on redshift $g(\mu_o,a_s)$, the inclination angle of 
the disk $\mu_o$, and the black hole spin $a_s$.
The magnification $A$ depends on $v_c$, $t_0$, $\phi_c$, $K$, and $A_0$, for a 
total of seven model parameters.

   We compute the transfer function by shooting rays from infinity
at a grid in $(\alpha, \beta)$ until they cross the equatorial plane.
The computational method is based on Rauch \& Blandford (1994), and is
described in Agol (1997).  We use a nested
grid of rays that is more finely sampled towards the center to resolve
the inner parts of the accretion disk in greater detail.  For a given
observed frequency $\nu_o$, we compute $\nu_e$ as well as the
emitted radius at each $(\alpha,\beta)$, and interpolate these on the
(pre-specified) grid of radii and frequencies at the 
accretion disk.

    To compute a normal transfer function, we would then simply sum over
the grid.  In this case, we multiply each ray by a further factor that
describes the magnification due to the microlensing at any particular time,
and then sum over the grid.  This procedure may be summarized
\begin{equation}\label{mateq}
F_i = T_{ij} I_j,
\end{equation}
where $i=1,N_F$ ($N_F= N_t\times N_\nu$ = \# observed frequencies $\times$ \# observed times)
labels each measured observed frequency/time, and $j=1,N_I$ 
($N_I=$ \# emitted frequencies $\times$ \# emitted radii) labels
each emitted frequency/radius pair.  The matrix ${\bf T}$
contains the integration and interpolation factors.

\subsection{Regularized Inversion}

Attempting to directly invert equation [\ref{mateq}] for ${\bf I}$ given
an observed set of ${\bf F}$ is impossible since the matrix ${\bf T}$
is generally singular, so that noise in the lightcurve is magnified strongly
during inversion.  This fact requires the introduction of some sort of
{\it a priori} knowledge in order to make inversion feasible.  Regularization
is a particularly useful way to do this, as discussed in Press et al. (1992), 
because
the ``prejudice" injected into the solution is usually relatively benign
and also relatively controllable.  The essence of the linear regularization method
is to minimize both the deviation of the model from the data, and also the
deviation of the model from ``smoothness," as defined by some sort of
differencing operator.  In our case, at any given frequency,
we expect the emitted intensity
$I_\nu$ to be smooth as a function of radius, but not the one-dimensional
profile $P_\nu$.  

     In our specific implementation of the method, we also impose several
other restrictions on the solution.  We expect that the emitted intensity in
the fluid frame diminishes as the black hole event horizon is approached;
we therefore require $I_\nu(r)$ to approach zero as $r$ approaches $r_g$.
In fact, because those regions are so strongly redshifted from almost
any inclination angle, the intensity in the fluid frame is almost
completely unconstrained by the data, so physical assumptions have a
very strong impact on the solution in this region.
Similarly, we also require $I_\nu$ to approach zero at very large radii, for
there is little energy available there to dissipate.  Particularly at
low frequency, it may sometimes be desirable to relax this constraint.
In addition, we would like the inverted intensities to be
positive definite; to achieve this, we maximize $A_I=(\sum_i I_i w_i)/N_I$,
where $w_i$ is a weighting factor.  The most appropriate weighting is
$w_i=r_i^2$ since the radiating area associated with each logarithmic radius
interval scales as $r_i^2$.  Finally, for any choice of grid, there
will always be some radius/emitted frequency pairs that are not constrained by
the data because Doppler shifts push $\nu_o$
outside the observed region (these are the intensities for which
the corresponding column of ${\bf T}$ is all zeros).  In order to prevent those
frequencies from contributing to the smoothing condition, we require
that the associated intensities be zero.

Combining all these considerations leads to the following
regularization operator:
\begin{eqnarray}
{\cal B} = {1\over A_I^2}\left\{
{\sum_{i=N_\nu+1}^{N_I} (I_i w_i-I_{i-N_\nu}w_{i-N_\nu})^2 +
N_I(I_1w_1)^2+ }\right. \cr
\left. {N_I(I_{N_I}w_{N_I})^2+N_I^4\sum_{\{j:{\rm if} T_{ij}=0 \forall i\}} (I_jw_j)^2
}\right\},
\end{eqnarray}
where $N_\nu$ is the number of frequency grid points at the
accretion disk.  The first sum describes the smoothness
as a function of radius at each frequency, the two isolated terms
give the boundary conditions at the innermost and outermost radii,
and the last sum is the factor encouraging minimization of those
$I_j$ unconstrained by data.  Note that the specific intensity vector
is ordered with all the frequencies at one radius grouped together,
so that $I_i$ and $I_{i-N_\nu}$ give the intensity at the same
frequency, but adjacent radii.
The smoothing operator ${\cal B}$ has the useful property of providing
a model-independent measure of the ``smoothness'' of different solutions,
due to the normalization by $A_I$.  Other differencing schemes might also
be used; in the examples we have explored, it makes little difference
to the outcome.

    Although the regularization condition is designed to be relatively
innocuous, no such injection of prejudice can be altogether free from
consequences (we will discuss the effect of our particular choice in
\S 3.2).  We stress that the details of the regularization condition
are always subject to ``tuning" in the light of either theoretical
expectations, or, better, the implications of real data.

To solve for ${\bf I}$, we minimize the following function:
\begin{equation}\label{fmin}
f({\bf I})={1\over N_F}\chi^2+\lambda{\cal B}
\end{equation}
with
\begin{equation}
\chi^2 = \sum_{i=1}^{N_F}\left({\sum_{j=1}^{N_I}T_{ij}I_j -  
F_i\over \sigma_i}\right)^2,
\end{equation}
where $\lambda$ is a constant, and $\sigma_i$ is the error on
$F_i$.
We start with a direct solution of $\partial f / \partial {\bf I} = 0$,
setting $A_I=1$,
\begin{equation}\label{isolve}
{\bf I} = \left({\bf M}^T{\bf M} + {\lambda\over A_I} {\bf H}\right)^{-1}{\bf M}^T {\bf G}
\equiv {\bf Q}{\bf M}^T {\bf G},
\end{equation}
where $M_{ij} = T_{ij} /\sigma_j$, $G_i = F_i /\sigma_i$, and
${\bf H}$ is defined such that ${\cal B} = {\bf I}\cdot{\bf H}\cdot{\bf I}$.
We then update $A_I$ from the solution, and iterate until
$A_I$ converges.  In some instances, $A_I$ can become negative
after an iteration.  If this happens, we re-set $A_I$ to be 0.1 times
its value at the previous iteration, and recalculate the step.

We assume that the magnification outside the caustic, $A_0$,
can be measured from the lightcurve, and subtracted off.  Then, the
parameter $K$ (the caustic magnification factor) is completely
degenerate with the disk surface brightness since a decrease in
magnification corresponds to an increase in the surface brightness
of the source.  Consequently, we can determine the {\it shape}
of the surface brightness profile, but not its absolute level.

Several other parameters also remain to be determined after
the direct inversion for the surface brightness profile.  We
call them collectively $\bzeta =(t_0, v_c, \mu, \phi_c, a_s)$.  To
find them, we fix $\lambda$ and compute $\chi^2$ over a coarse grid
in this five-dimensional parameter space.  Starting from the
$\bzeta$ giving the smallest $\chi^2$ in this grid, we refine our
estimate of these parameters using the Levenberg-Marquardt method.
We compute the partial derivatives of $\chi^2$ with respect to
the axes in $\bzeta$ space by
finite differences.   If a parameter with boundaries goes out of
bounds, we fix it at the value where it went out of bounds, and
keep it fixed throughout the rest of the minimization.  This generally
occurred with $a_s$ when it was near 0 or 1, and for $\mu$ and $\phi_c$
when the disk was face-on.
Fixing this improved estimate for the best-fit $\bzeta$, we increase
$\lambda$ and re-solve for the surface brightness profile
until $\chi^2 = N_F$.  

   The number of degrees of freedom against which to compare $\chi^2$
is not clearly defined for several reasons.  One is that many of
the model parameters are not entirely free; for several ($\mu$,
$M_{BH}$, $v_c$) there are prejudices or constraints from other
experiments.  Another reason is the variable weight given the
smoothing constraint {\it vis-a-vis} the data, as we are minimizing
$\chi^2 + \lambda {\cal B}$ rather than $\chi^2$.  In the limit of
large $\lambda$, there is effectively only one free parameter for
each frequency in the fit to the $I_i$; in the limit of $\lambda = 0$,
there are as many free parameters as there are grid points.
Given these considerations, $N_F$ is an upper bound to the true
number of degrees of freedom; by raising $\lambda$ until $\chi^2=N_F$,
we ensure that we do not overfit the data.

\subsection{Errors} \label{errors}% see pp. 95-96 of 11/18/98 notes

  The word ``error" has several different meanings in this context, and
it is important to distinguish them.  First of all, the errors in the
inferred intensities have different properties from the errors in the
model parameters.  Second, both are potentially subject to systematic 
error as well as random error.

    We will begin by estimating the random error in
the intensities ${\bf I}$.  Formally, we may say that
\begin{equation}\label{erri}
\delta I_i^2 = \sum_j\left({\partial I_i \over \partial F_j}\right)^2 \sigma_j^2,
\end{equation}
where ${\partial I_i \over \partial F_j}= \sum_k Q_{ik} T_{kj}/\sigma_j^2 $
(see equation [\ref{isolve}]), and we assume the fluxes have uncorrelated
errors.
The $\lambda = 0$ case is of special interest because it reveals
which $I_i$ (i.e., which frequency/radius pairs) are so constrained
by the data that even without regularization they may be reliably
determined.  In this
limit, ${\partial I_i \over \partial F_j} = T^{-1}_{ij}$, so the
uncertainty in $I_i$ is given by $\delta I_i^2 = W^{-1}_{ii}$
where $W_{ij}= \sum_k T_{ki}T_{kj}/(\sigma_i \sigma_j)$.
${\bf W}^{-1}$ can be computed by singular value decomposition; in
practice, we replace the singular values with a small number.
We show an example of this procedure in Figure 9.

The ``formal accuracy" of our inversion is illustrated by a plot
of $U_i \equiv I^r_i/\delta I_i$ (the superscript $r$
stands for recovered intensity) as a function of frequency
and radius, computed from equation [\ref{erri}] (Figure 9).
The results for both $\lambda =0$ and the
maximum $\lambda$ consistent with the data are shown.  In the case
$\lambda = 0$, we use the original $I_i$ instead of the recovered values
to compute $U_i$.
The formal accuracy depends on the true surface brightness: for a given 
radius, $U_i$ tends to peak where the flux is largest.  Also,
$U_i$ diminishes at large radii, since those radii aren't monitored
for long enough to truly determine $I_i$.

In real solutions $\lambda \neq 0$, and the smoothing operator correlates
the intensities at neighboring radii sharing the same frequency.
The uncertainties in this case are most easily
estimated by a Monte Carlo procedure in which the lightcurves are
perturbed by random realizations of noise in the data.  The
distribution of $I_i$ after re-solving each of these realizations
gives the random error in $I_i$.  In evaluating these estimates,
it is important to understand that points weakly constrained by the
data have little sensitivity to measurement errors because they
are primarily determined by the smoothness constraint.  As a result, 
their random errors are artificially small.

   To check for systematic errors in the intensities, we will compute
the difference between the original and recovered surface brightness,
$\Delta I_i = I^o_i - I^r_i$, where superscript $o$ stands for ``original." 
The systematic error is, of course, far more strongly model-dependent than
the random error.  It depends on the real intensity distribution, the
character of the data (particularly the sampling), and the inversion scheme.
These considerations will be discussed at greater length in 
\S  \ref{bestcase}.

    We estimate the uncertainties in the model parameters
$\bzeta$ two ways: through the same Monte Carlo procedure as for the
$I_i$, and also through mapping out the $\chi^2$ found by direct
solution of the original data (at fixed $\lambda$) for different
choices of $\bzeta$.  Examples will be shown in \S \ref{bestcase},
\ref{varyparam}.

\section{Simulations}

\subsection{Range of Parameters Examined}

To determine how the inversion works in practice, we performed
simulated inversions, varying the parameters describing the underlying
model, the parameters describing the data set, and the parameters
specifying the details of the solution technique.  By varying
the model parameters, we learn about whether the method is
sensitive to the intrinsic nature of the quasar, or the microlensing
event; by varying the observational parameters, we determine
what the requirements will be for successful experiments; by
varying the solution parameters, we learn how to tune the solution
technique for optimum results.

\subsubsection{Model parameters $\bzeta$}

   In all of our simulations we assume that the intrinsic radiated
intensity in the fluid frame is a black body at the local effective
temperature, isotropic in the outer half-space.  Detailed non-LTE
spectra computed for our fiducial parameters (see the next several
paragraphs) are consistent with both the observed optical/ultraviolet
spectrum and the microlensing size constraint for the Einstein Cross, 
given the freedom to choose an extinction correction and macrolens 
magnification (Hubeny \& Agol, in preparation).
However, we examined a number of possibilities for the other
parameters defining the intrinsic character of the quasar and the
microlensing events.

Although we do not know the mass of the black hole, we may set
reasonable bounds on what it could be. 
If the intrinsic bolometric luminosity is $3\times 10^{46}$ erg/s
(Rauch \& Blandford 1991), the quasar would be at its Eddington limit
if $M_{BH}\sim 2\times 10^8 M_\odot$.  On the other hand, the size of
the optical emitting region is limited to no more than
$\sim 2\times 10^{15}$~cm at $\sim 10^{15}$ Hz (quasar rest frame) from
microlensing (Wambsganss et al. 1990).  If this equals the radius of
maximum emission for an accretion disk ($\sim 10 r_g$), then
$M_{BH}\sim 10^9 M_\odot$.  On this basis we suppose that the true
mass is between $2\times 10^8 M_\odot$ and $10^9 M_\odot$.  Our
choice for the fiducial model will be $10^9 M_{\odot}$.

The units we used in the simulations are $r_g$ for length
and $\Delta t$, the sampling rate, for time.  The units of $v_c$
are then $r_g/\Delta t$:
\begin{equation} %from p. 82 of OUX microlensing notes
v_c = 0.29 \left({V_c \over 5000 {\rm km/s}}\right)\left({M_{BH}\over
10^9 M_\odot}\right)^{-1} \left({\Delta t \over 1 {\rm day}}\right),
\end{equation}
where $V_c$ is the caustic velocity in km/s with distance measured
at the quasar plane, while time is measured at the observer.
The caustic velocity, $V_c$, is quite uncertain, but is likely
to be in the range $3000 - 5000$ km/s (Wyithe et al. 1999). 
For $\Delta t=3$ days, this
corresponds to $0.5 \leq v_c \leq 4$, the range we span in our
simulations.  In the fiducial model, we choose $v_c = 1$.

We try values of $\phi_c$ between $0$ and $2\pi$,
with $\pi/2$ for our fiducial model.

We choose $t_0=0$ for our
fiducial model, and also try $t_0=5$ to see whether this technique
works when the central time of the monitoring does not coincide
with passage of the caustic line across the center of the black hole.

There are no observational estimates of the inclination angle;
however, unification arguments 
for radio-loud AGN suggest that quasars are less face-on than blazars, 
but closer to face-on than radio galaxies, so we choose a fiducial
inclination of $30^\circ$.  We also look at cases with $\theta = 0$
and $\theta = 60^{\circ}$.

Because accretion can spin up black holes,
and because Kerr holes permit more efficient accretion than Schwarzschild
black holes, we choose $a_s=0.998$ for our fiducial model, but also
study one example with $a = 0$.

The last parameter is the accretion rate.  With the fiducial choices
for the other parameters, the observed spectrum is best reproduced
with $\dot m \approx 1 = \dot M/(1 M_\odot/yr)$.

\subsubsection{Observational parameters}

We vary the number of observations, time sampling interval, SNR,
and number of observed wave bands, as well as the model parameters.
It is especially important to determine how the quality of the
result depends on the number of observations 
because these observations must be targets of opportunity, and
thus will impact other observations at a given telescope.
We explore what happens for experiments
with between 5 and 41 observations (in all cases, we assume
uniform spacing).  The ratio between $\Delta t$ and the duration
of the microlensing event is implicitly given by $v_c$.

We try two choices for the SNR (as measured outside the microlensing
event): 100 for each image in the best case, and 50 in the worst case
(these were chosen based on current ground-based errors, Rachel
Webster, priv. comm.).

Since disks are broad-band emitters, a broad range of observing
frequencies is necessary.  Observations in the four wave bands
V, B, R, and I (or equivalent) should be routine; observations
in U, J, H, K, or in the UV with HST will be much more difficult
to obtain, but will yield much more information.  To see just
how important the additional bands are, we try using just
ground-based data in 4 or 8 bands, or 8 ground + 3 HST bands in the
best case.  The short-wavelength bands are especially important
for hot disks, since the deepest part of the potential well is
seen at the shortest wavelengths. 

\subsubsection{Solution parameters}

    Several considerations determine the number of frequency and
radius points at which we may solve for the surface brightness.
The number of frequency points is not simply equal to
the number of colors at which the quasar is monitored because of
the extensive Doppler shifting.  We found that in practice the
best solution grids in both frequency and radius space were
logarithmic.  We solve for the intensity at frequencies
equally spaced logarithmically between $3\times 10^{14}$~Hz and
$10^{16}$~Hz, and radii equally spaced logarithmically from $r_g$ to
$r_{out}$, with $r_{out}=500r_g$. 
In the initial testing of the inversion using the fiducial model, we
found the smallest number of radii and emitted frequencies for which
we could obtain $\chi^2 = N_F$ for some $\lambda$ was
15 radii and 10 frequencies, which we subsequently used for all the
simulations.

\subsection{Best-case simulation} \label{bestcase}

  For our ``best-case'' simulation (designated A1 in Table~1),
we fixed $\bzeta$ at the fiducial
parameter choices.  The observational parameters were: 41 observations,
$SNR = 100$, and 11 spectral bands.  

\vskip 2mm
\hbox{~}
\centerline{\psfig{file=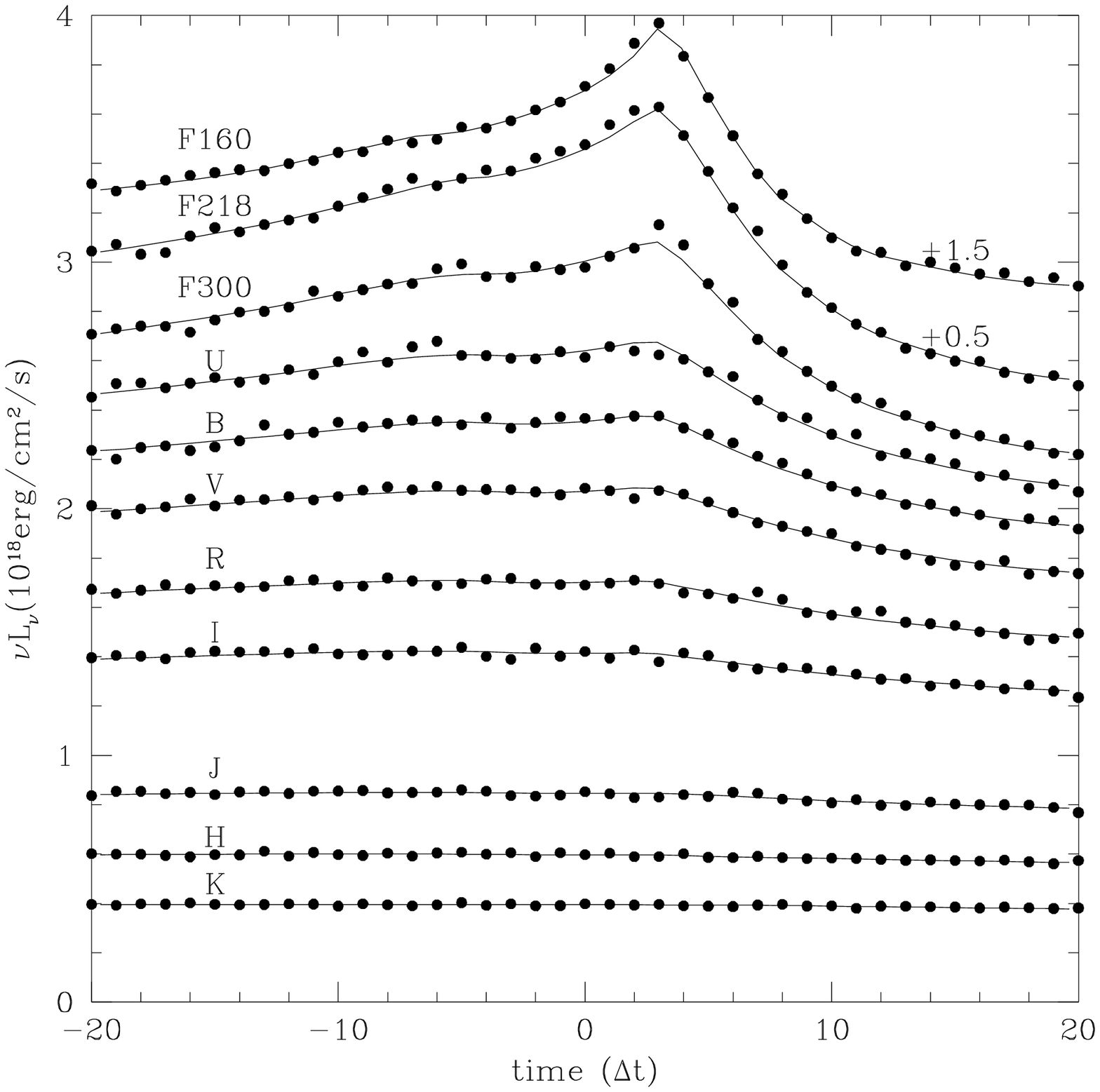,width=3.7in}}
\noindent{
\scriptsize \addtolength{\baselineskip}{-3pt}
\vskip 1mm
\begin{normalsize}
Fig.~5.\ Solid points are lightcurves with noise added. The top two curves
have been shifted upwards by the amount indicated for clarity. The solid
lines are the lightcurves from the reconstructed disk profile. %\label{fig05}
\end{normalsize}
\vskip 3mm
\addtolength{\baselineskip}{3pt}
}
Figure 5 shows the lightcurves for this example,
with the observed bands de-redshifted for $z=1.695$.
Note that the higher
frequencies, which come from nearer to the black hole, are magnified
more strongly than the lower frequencies.
Figure 6 shows the original and reconstructed disk
intensity, $I(\nu_e,r_e)$, as a function of frequency and radius
for the best fit parameters for this best case (see Table 1).  
The overall shapes are reproduced quite well.  Figure 7 shows
the same results in a different format: we have multiplied surface 
brightness times $r_{e}^2$, and plotted the data differently for clarity.
Note that the low and high frequencies and small and large radii
are poorly constrained since the simulated lightcurve only covers
$-20 r_g < r < 20 r_g$ and 1600~${\rm \AA} < \lambda < 3~\mu$m. 
Consequently, at these points the regularization tries to make the
flux per log radius constant as a function of radius.
Figure 8 shows the ratio of the reconstructed to the original 
one-dimensional profile,
$P_{\nu_o}(x)$ (computed from $I^r_i$) for run A1.
\vskip 2mm
\hbox{~}
\centerline{\psfig{file=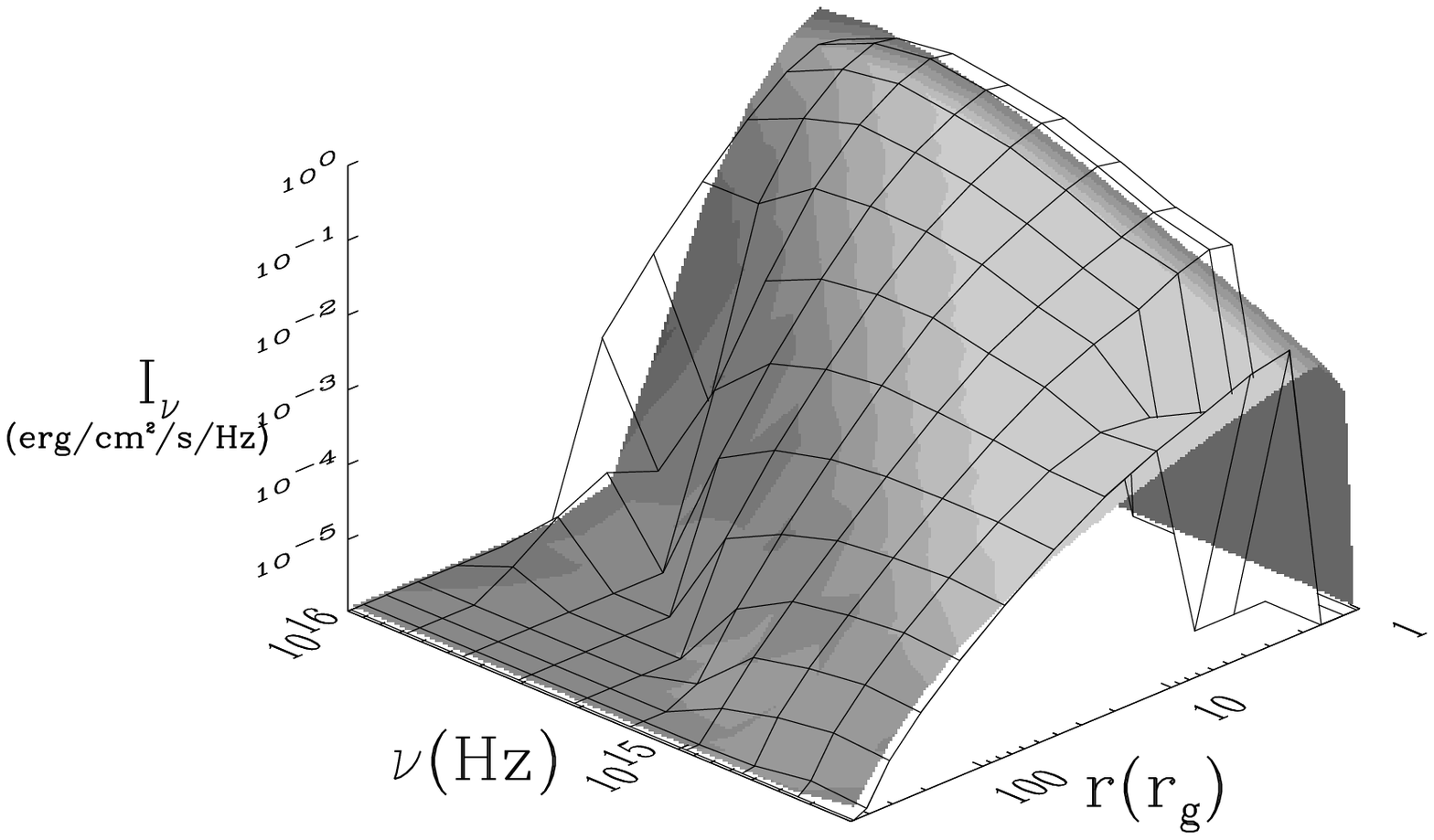,width=3.5in}}
\noindent{
\scriptsize \addtolength{\baselineskip}{-3pt}
\vskip 1mm
\begin{normalsize}
Fig.~6.\ Specific intensity $I_\nu(r_e)$ versus emitted frequency
and radius.  The shaded surface is the original surface brightness,
while the skeleton plot is the surface brightness recovered.
\end{normalsize}
\vskip 3mm
\addtolength{\baselineskip}{3pt}
}
\vskip 2mm
\hbox{~}
\centerline{\psfig{file=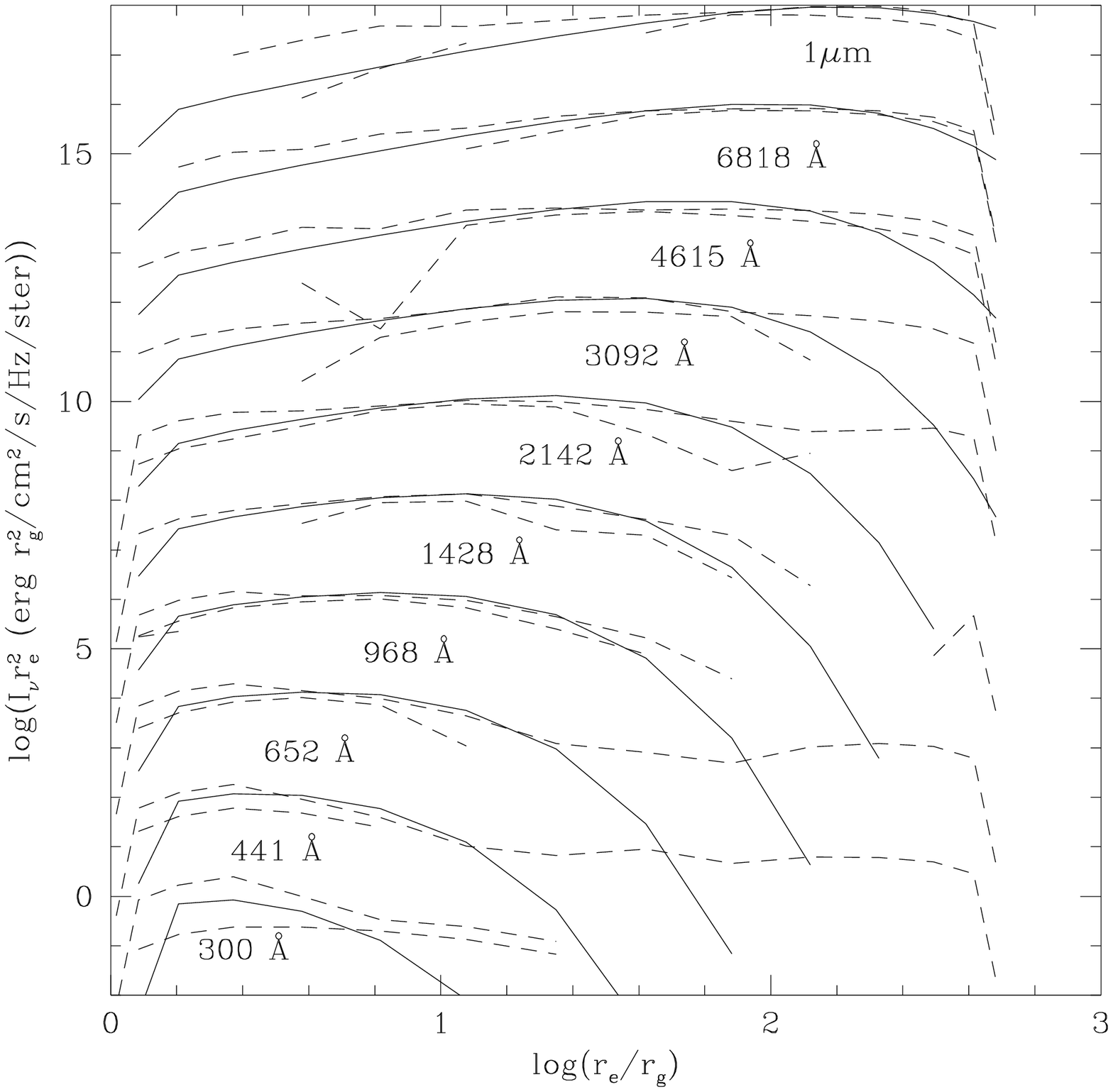,width=3.7in}}
\noindent{
\scriptsize \addtolength{\baselineskip}{-3pt}
\vskip 1mm
\begin{normalsize}
Fig.~7.\ Specific intensity $I_\nu(r_e)$ times $r_e^2$
versus $r_e$ for each emitted frequency. The solid curves are the
original $I_\nu r_e^2$, while the dashed curves are the minimum and maximum
values of the recovered $I_\nu r_e^2$ for 20 simulations.  Each curve is
shifted upwards by 2 with respect to the curve below - the zero point
is for the lowest curve.  We have not plotted negative intensities.
\end{normalsize}
\vskip 3mm
\addtolength{\baselineskip}{3pt}
}
\vskip 2mm
\hbox{~}
\centerline{\psfig{file=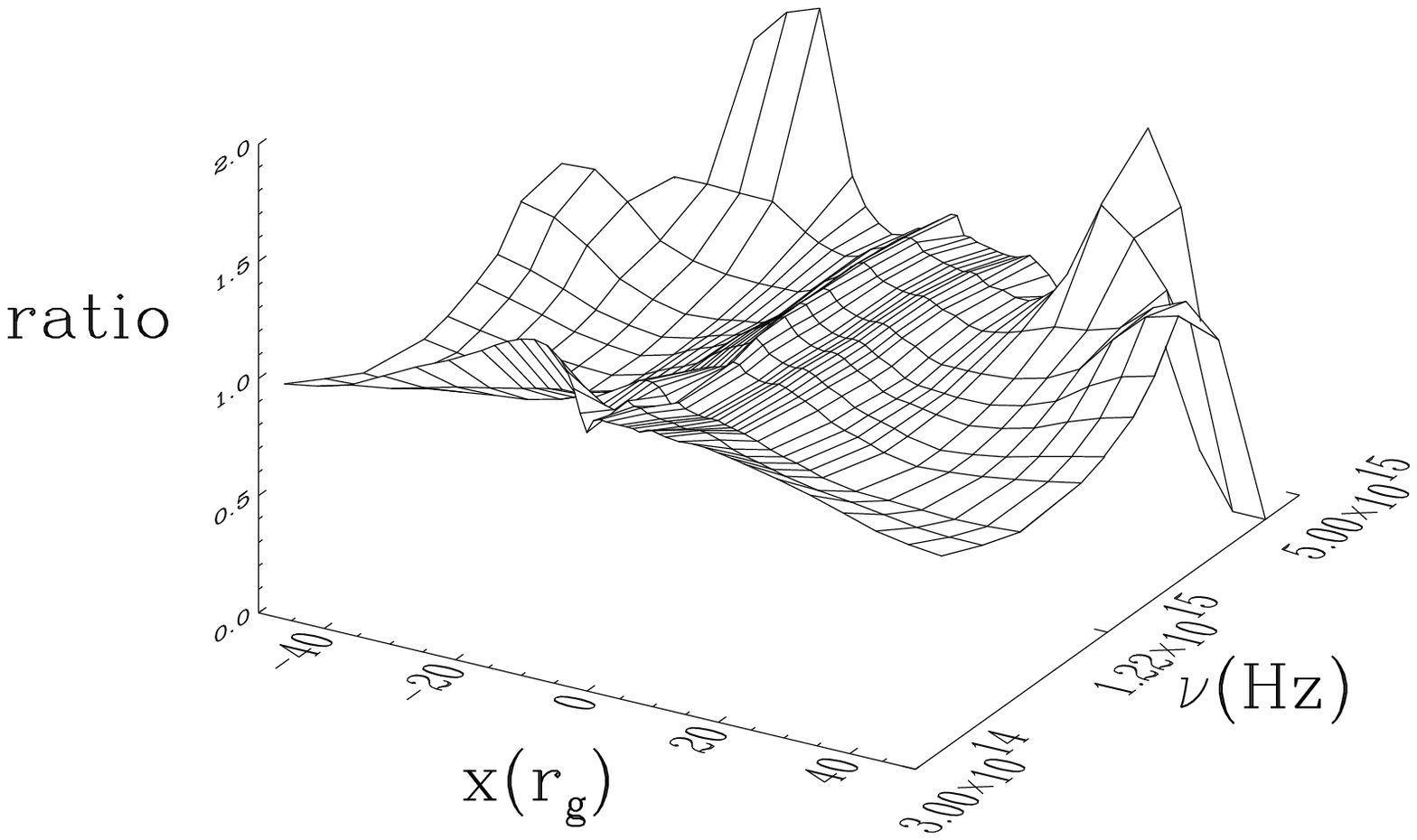,width=3.7in}}
\noindent{
\scriptsize \addtolength{\baselineskip}{-3pt}
\vskip 1mm
\begin{normalsize}
Fig.~8.\ Ratio of recovered to original one dimensional disk profile as a
function of position and observed frequency.
\end{normalsize}
\vskip 3mm
\addtolength{\baselineskip}{3pt}
}

   To discuss the reliability of this solution, we begin by contrasting the
region in the $r_e$--$\nu_e$ plane where the random error is predicted
to be relatively small with the region where the actual error is small.
As can be seen in Figure 9, the region of large $U_i$ 
for $\lambda \neq 0$ largely,
but not entirely, coincides with the region of large $I_i^o /|\Delta I_i|$.
Moreover, both of these regions follow a track defined, not surprisingly,
by the requirement that $r_e^2 I_{\nu_e}(r_e)$ is relatively large.  Elsewhere
in the plane, the contribution to the flux is so small that the intensity
is virtually unconstrained by the data.
\vskip 2mm
\hbox{~}
\centerline{\psfig{file=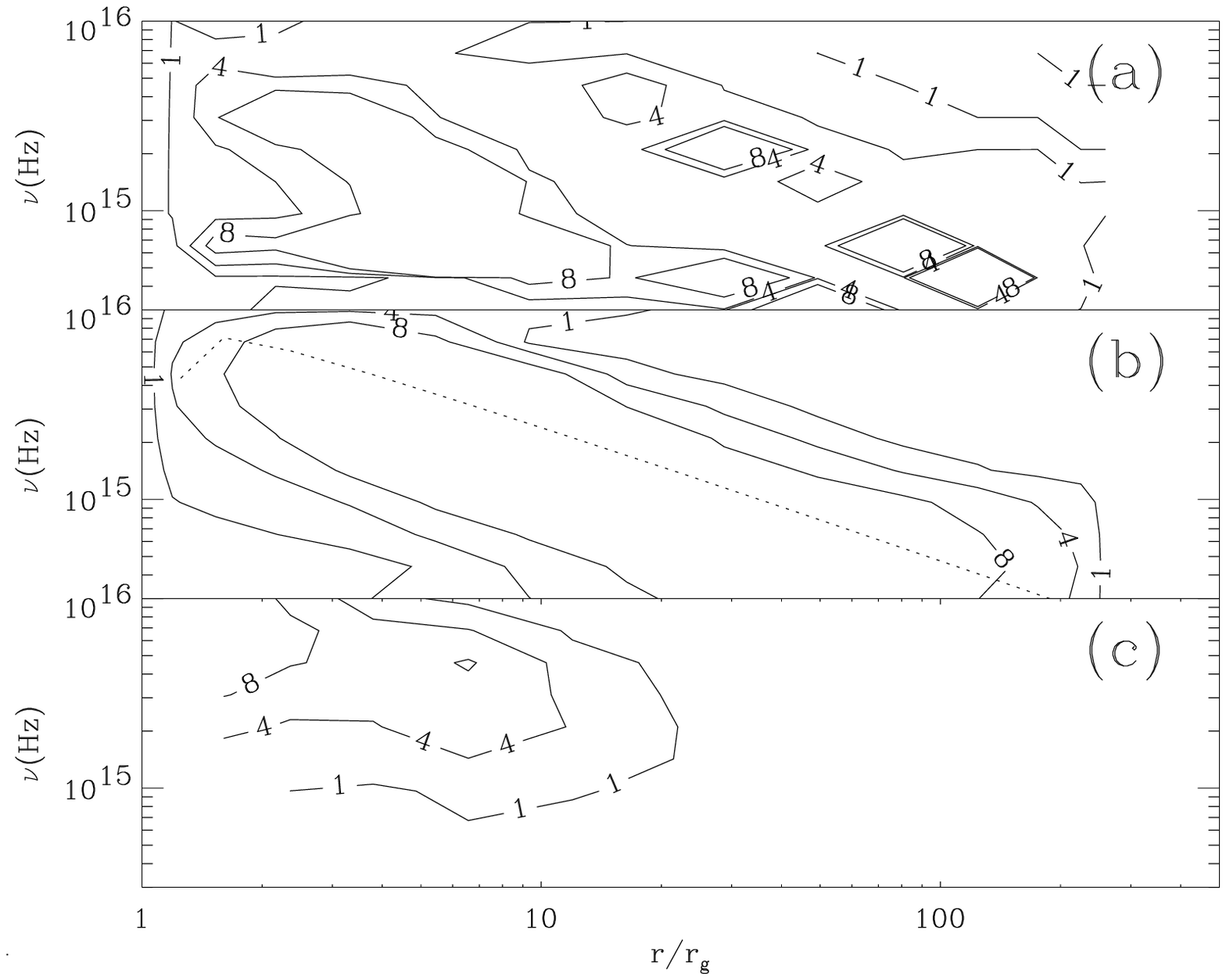,width=3.7in}}
\noindent{
\scriptsize \addtolength{\baselineskip}{-3pt}
\vskip 1mm
\begin{normalsize}
Fig.~9.\ Plot of $I^r_i/\Delta_i$ (a) and formal accuracy, $U_i$, of
reconstructed disk profile for $\lambda = 1.1$ (b) and $\lambda=0$ (c).  
The dotted line in panel (b) shows where the peak of $B_\nu$
occurs at each radius.
\end{normalsize}
\vskip 3mm
\addtolength{\baselineskip}{3pt}
}

    Where $U_i \simeq I_i^o/|\Delta I_i|$, the error is predominantly random
error, and $\delta I_i$ is a good predictor of its magnitude (in fact, in
this region $\Delta I_i$ has a Gaussian distribution of the correct
width).  However, there is also a zone on the large radius
side of the high-intensity track where the systematic error is as large or
larger than the random error.  The nature of this systematic error is
revealed by studying Figure 7.  The smoothing condition tends to
raise the intensity in regions where it should be small, and diminish it
where it is large.  Because $U_i$ is rarely large enough to be interesting
where $I_i$ is small, it is the latter effect that dominates in the region
of the $r_e$--$\nu_e$ plane highlighted in figure 9.  At least
within the context of this model, this systematic error is not the
result of the specific choice of smoothing constraint:  we
have tried a second-order linearization scheme to see if we could get rid
of the systematic deviation; however, we still found that the recovered
intensity was flatter than the original.  If we relax the condition that
the intensity should be zero at the last radial bin, then the intensity
approaches a constant for each frequency at large radius.

\vskip 2mm
\hbox{~}
\centerline{\psfig{file=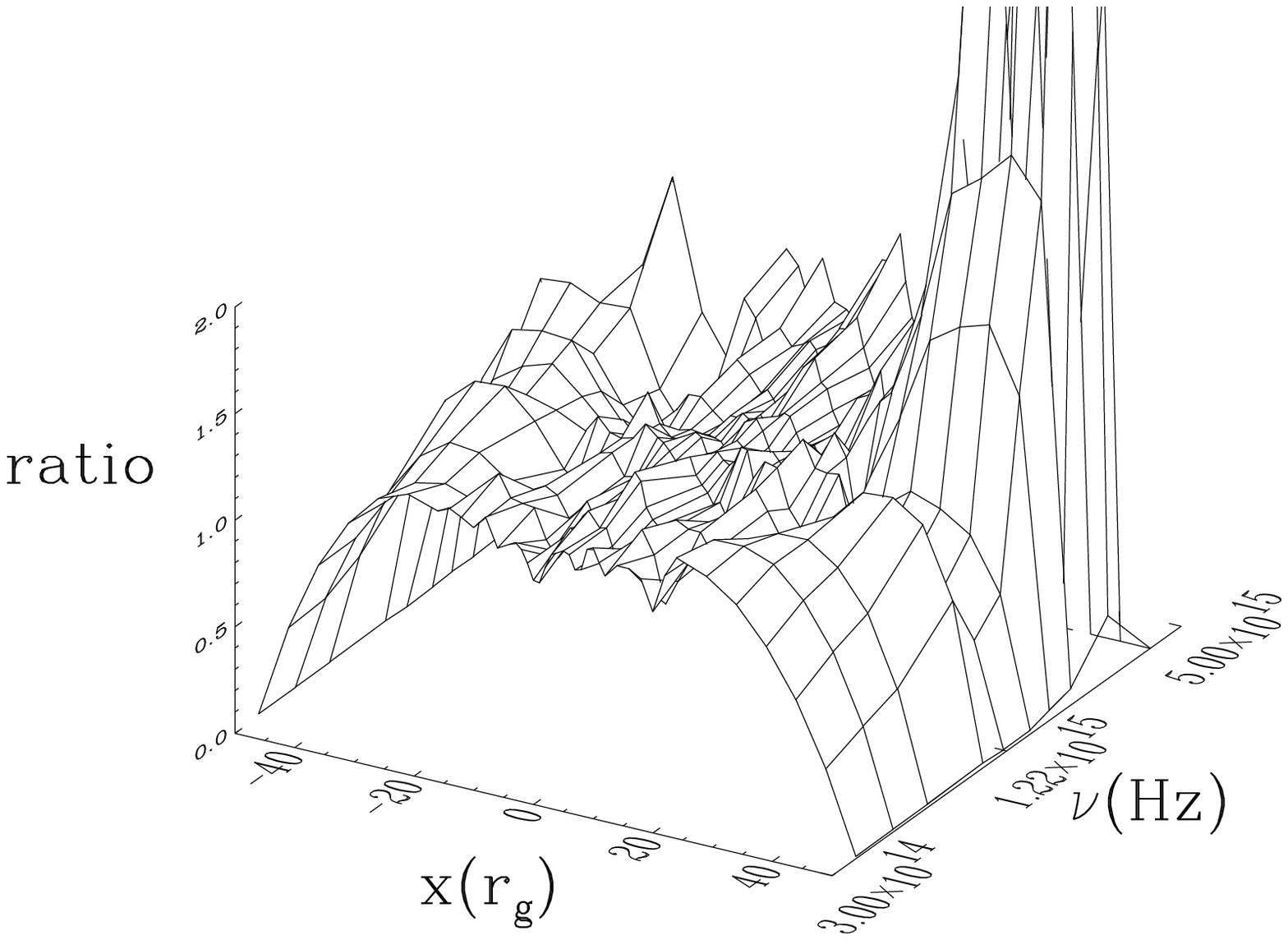,width=3.7in}}
\noindent{
\scriptsize \addtolength{\baselineskip}{-3pt}
\vskip 1mm
\begin{normalsize}
Fig.~10.\ Ratio of recovered to original one dimensional disk profile
as a function of position and observed frequency, using the technique
of Grieger et al. (1990).%}\label{fig7}
\end{normalsize}
\vskip 3mm
\addtolength{\baselineskip}{3pt}
}

We performed a regularized inversion using the Grieger et al. (1991)
technique for comparison.
Figure 10 shows their inversion on a data set equivalent to run
A1.  Since the regularization constraint attempts
to smooth $P_\nu$, the Doppler peaks are smoothed over, and the noise from the
lightcurve appears to still be present in the $P_\nu$.

After maximizing $\lambda$ consistent with $\chi^2 = N_F$, we then
fix $\lambda$ and vary each component of $\bzeta$, minimizing $\chi^2$
with respect to the other parameters.  This procedure shows how well
each parameter can be constrained for a given simulation, or whether
there are other local minima.  In Figure 11 we show the
$\Delta\chi^2$ for each model parameter.  For this particular
model, the physically interesting parameters, inclination angle ($\mu$) 
and caustic velocity ($v_c$), have well-defined minima.  The time of
origin crossing ($t_0$) and the caustic crossing angle ($\phi_c$) are
also well-behaved.
The black hole spin has a rather flat $\Delta\chi^2$ distribution.
However, the minimum does lie at the correct value.  Figure 11 also
has a histogram of the parameters from each noise realization,
showing that the minimum of the $\chi^2$ distribution has few outliers.
\hbox{~}
\centerline{\psfig{file=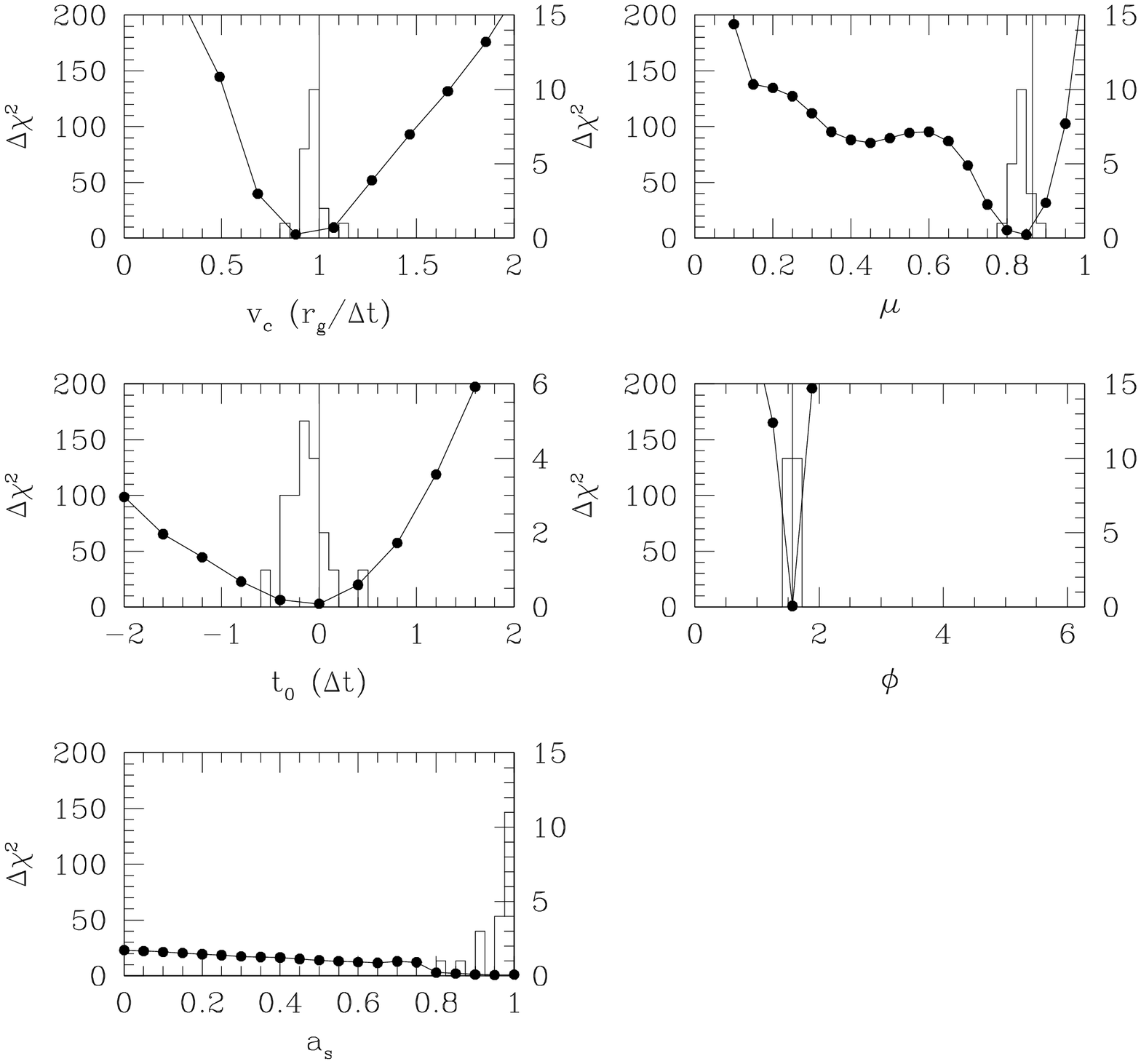,width=3.7in}}
\noindent{
\scriptsize \addtolength{\baselineskip}{-3pt}
\vskip 1mm
\begin{normalsize}
Fig.~11.\ Change in $\chi^2$ versus each parameter for the A1 case.
The vertical lines show the values of the original parameters.
The histograms show the best-fit parameter results of 20 lightcurve
realizations, with the right hand axis labelling the number in each bin.
\end{normalsize}
\vskip 3mm
\addtolength{\baselineskip}{3pt}
}

\subsection{Varying model parameters} \label{varyparam}

Table 1 shows the results of varying the model parameters,
keeping the observational parameters fixed at the ``best-case'' values.

Runs A1-A10 each have recovered surface brightness which
look similar to A1, and the $U_i$ are quite similar. 
Runs A11a and A12 reproduce the intensities well
for the lowest frequencies and for radii outside $r_{ms}$; however, the
recovered intensities are non-zero inside $r_{ms}$, contrary to the input model.

Table 1 shows the average and standard deviation of the 
recovered parameters, $\bzeta$, measured for 20 Monte Carlo realizations.
In all the cases we examined, the distribution of recovered parameters
is centered near the true model parameters, showing that there are no
systematic offsets introduced by our inversion.  This is encouraging
since it means that this technique has the potential to measure
important global properties of the accretion disk/black hole system.

Run A2 shows that we can determine the time that the caustic crosses
the black hole rather accurately.  We have also tried cases with
$t_0=\pm 10$, and we find that these are also measured quite well,
and that the intensity is reproduced as well as in the A1 run.

Runs A3 and A4 show that we can
distinguish between different caustic velocities, which means that 
we can constrain the black hole mass in terms of the lens velocity.
Runs A5 and A6 show that we can measure the disk inclination angle
for a wide range of intrinsic angles.  
Runs A7, A8, A9, and A10 show that we 
can measure the angle at which the caustic crosses
the accretion disk rather accurately.  
In some cases there is a degeneracy between $\phi_c$ and $2\pi-\phi_c$ 
when the disk inclination is small, but this should not affect the recovered 
surface brightness since the disk is approximately symmetric in this case.

Figure 12 shows the $\chi^2$ topology for each parameter for Run 
A10, a somewhat special case in which $\phi_c=\pi$.  The parameters
$t_0$ and $\mu$ have local minima away from the correct minimum;
however, these can be ruled out because some inferred intensities
have large negative excursions in the false minimum.
\vskip 1mm
\hbox{~}
\centerline{\psfig{file=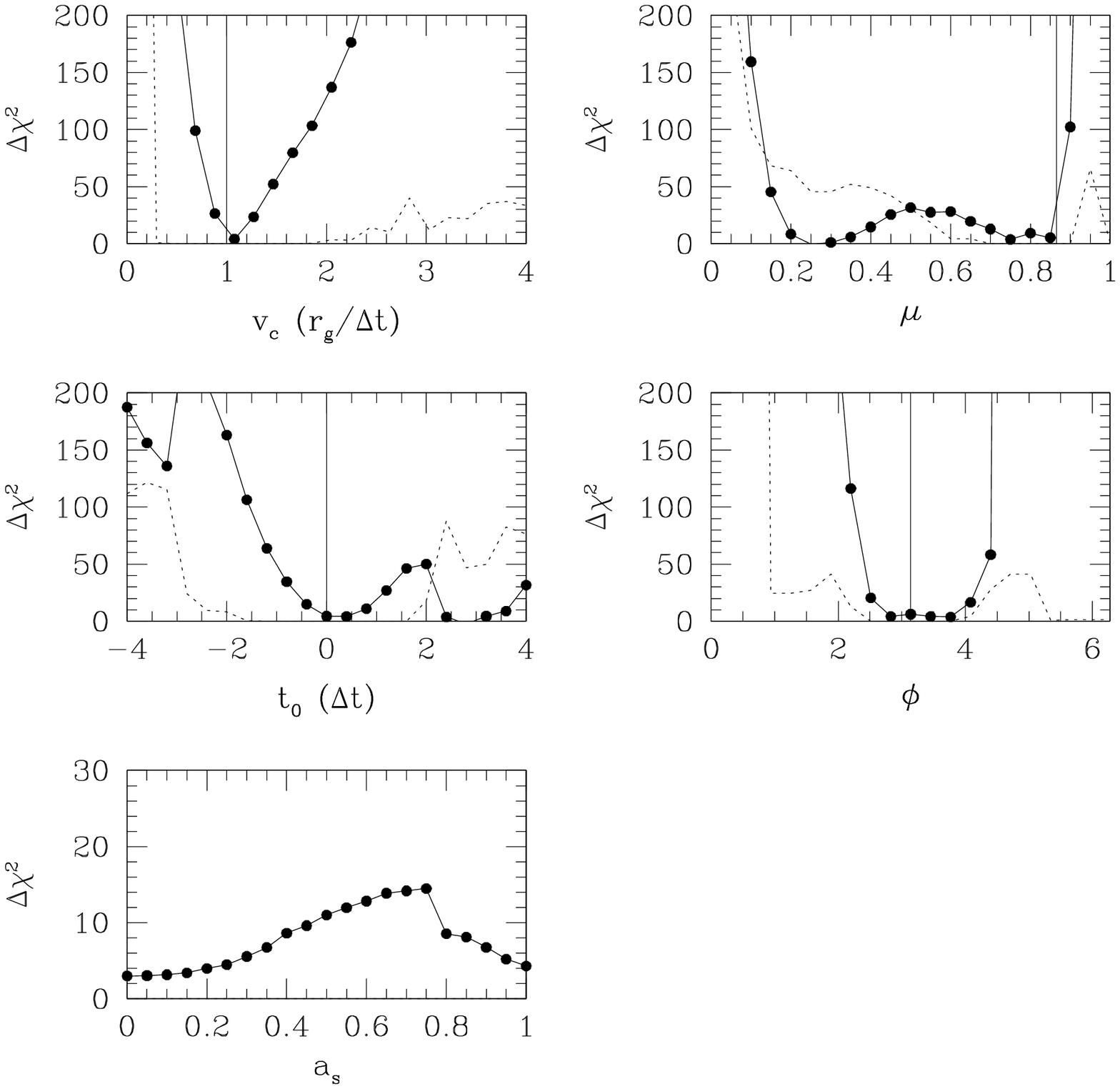,width=3.7in}}
\noindent{
\scriptsize \addtolength{\baselineskip}{-3pt}
\vskip 1mm
\begin{normalsize}
Fig.~12.\ Plot of $\Delta\chi^2$ (minimized over all other parameters) vs. each
parameter for run A10.  The solid vertical lines show the original parameters.
The dotted lines show the absolute value of the sum of the inverted
intensities which are negative (the scale is from 0 to 1).
\end{normalsize}
\vskip 3mm
\addtolength{\baselineskip}{3pt}
}

 Constraining the spin can be difficult, particularly when
$\phi_c \simeq \pi$.  For example, in Run A10 $a_s$ is not constrained 
at all (see Figure 12).  $\chi^2$ has as deep a minimum 
at $a_s = 0$ as it does at the
correct value $a_s = 0.998$.  Only if the coarse search in
$\bzeta$ space is lucky enough to discover the true minimum
will the Levenberg-Marquardt procedure home in on the correct value.
It is not clear why some $\phi_c$ are more favorable for determining
$a_s$.

  Whether $a_s$ can be constrained at all
depends on how strongly one believes in the model.  If no emission
is permitted inside the marginally stable orbit (Run A11b), $a_s$
can be constrained because there is a sizable difference between the
marginally stable orbit around a Schwarzschild black hole ($6 r_g$)
and a maximal Kerr black hole ($\simeq r_g$).  However, if one
is unwilling to make this assumption, the distinction between the
spins largely disappears (Runs A11a and A12).  The reason for this
indistinguishability is shown in Figure 13, which shows
a contour plot for the redshift as a function of 
position for black holes with spins $a_s=0.01$ and $a_s = 0.99$,
including the regions inside $r_{ms}$.  The two plots are almost
identical around $10 r_g$, where most of the observed
radiation comes from in this model.  Figure 14
shows the $\chi^2$ (minimized over all other parameters) vs. $a_s$
and recovered paramters for 25 simulations assuming that emission only 
occurs outside of $r_{ms}$.  The $\chi^2$ has a clear minimum near
the correct spin, and the simulations show that the spin can be
rather accurately recovered.  Since $r_{ms}$ increases with decreasing
spin, we can only hope to obtain a lower limit on $a_s$ by assuming
$r > r_{ms}$.  Indeed, the $\chi^2$ vs. $a_s$ is flat in the case
of zero spin (A11a).
\vskip 2mm
\hbox{~}
\centerline{\psfig{file=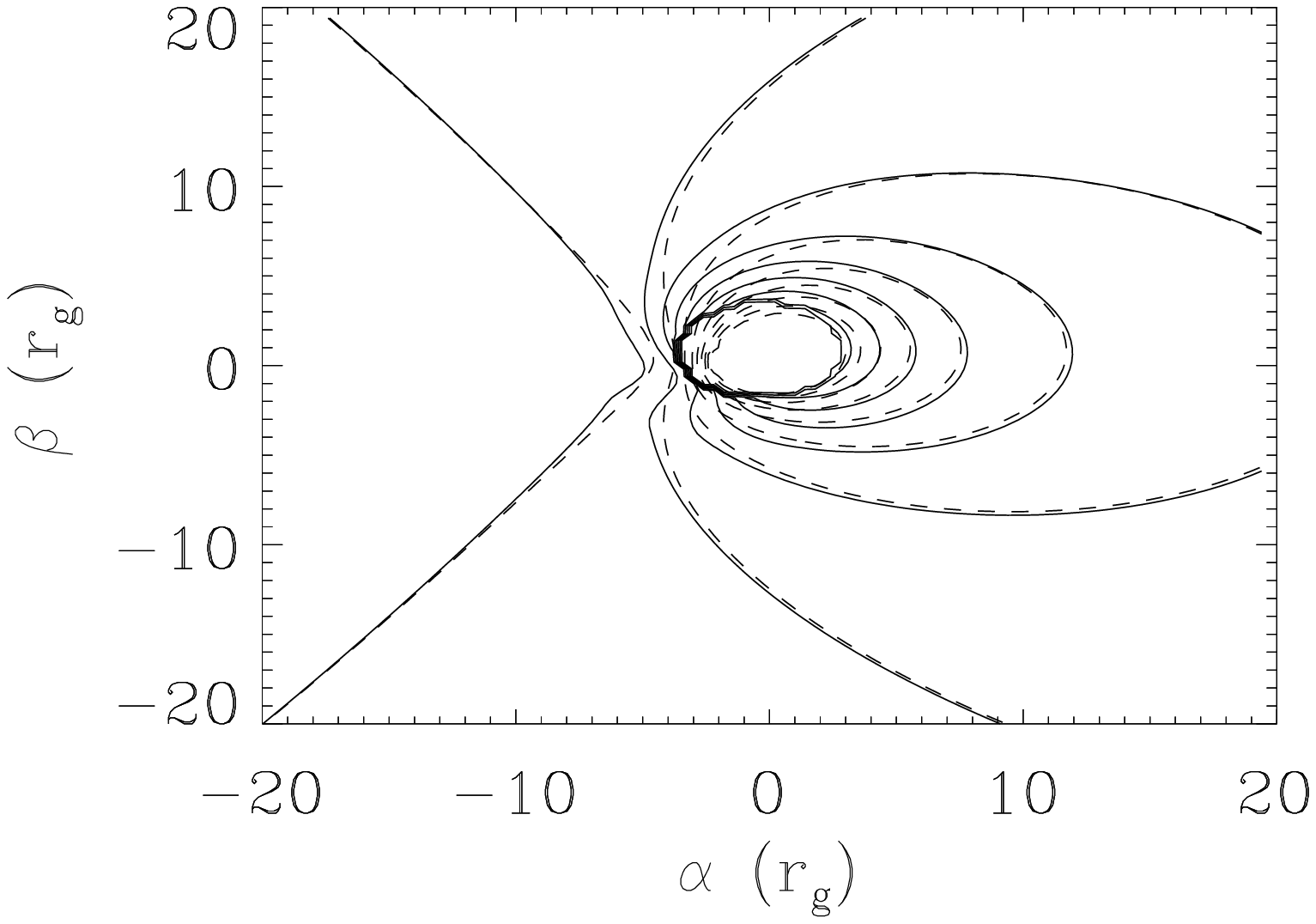,width=3.7in}}
\noindent{
\scriptsize \addtolength{\baselineskip}{-3pt}
\vskip 1mm
\begin{normalsize}
Fig.~13.\ Redshift as a function of position for black holes with spin 
$a_s=0.01$ (solid line) and $a_s=0.99$ (dashed line).
\end{normalsize}
\addtolength{\baselineskip}{3pt}
}
\hbox{~}
\centerline{\psfig{file=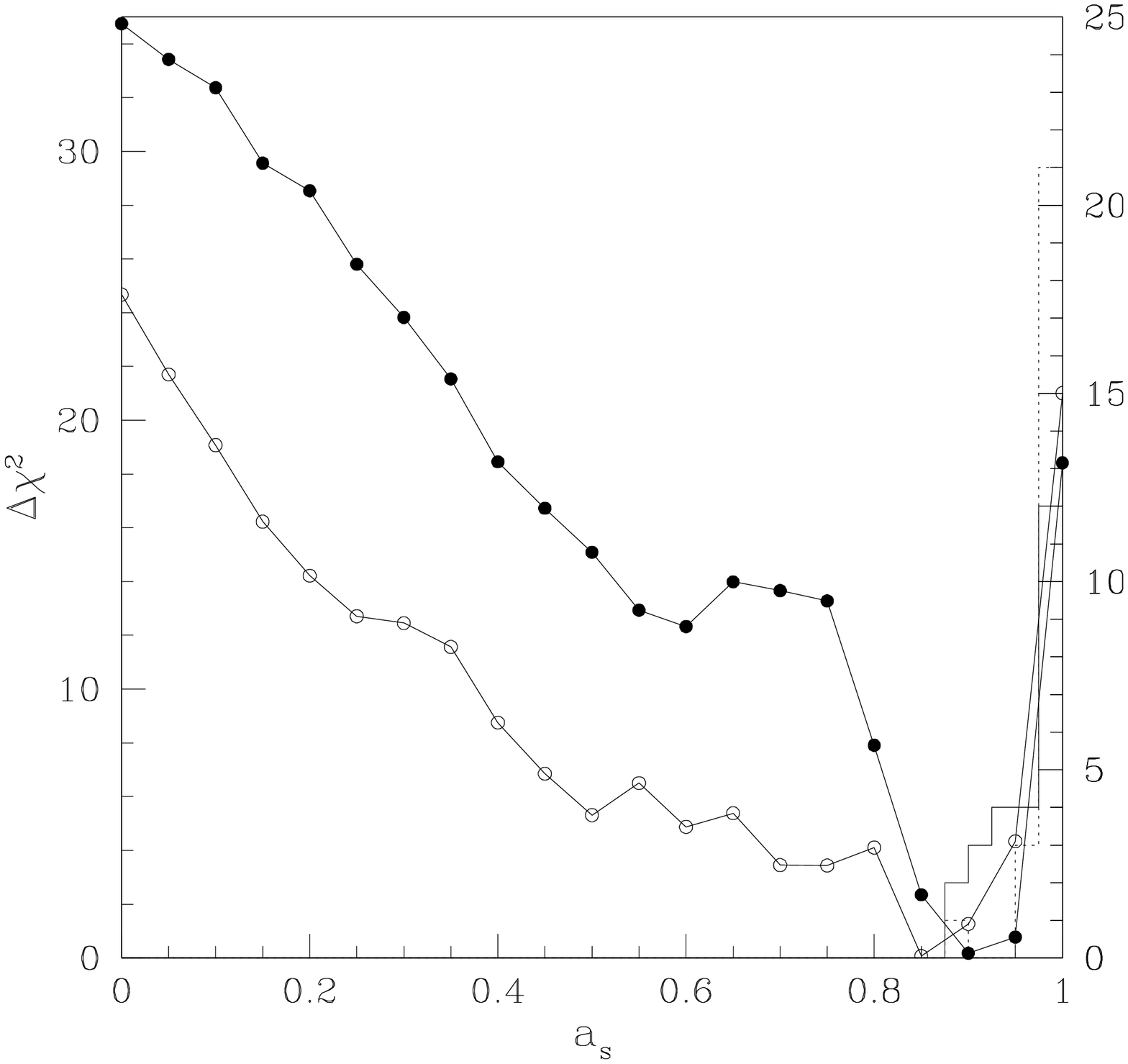,width=3.7in}}
\noindent{
\scriptsize \addtolength{\baselineskip}{-3pt}
\vskip 1mm
\begin{normalsize}
Fig.~14.\ Plot of the $\chi^2$ vs. $a_s$ (minimized over all other parameters)
for cases A1 and A10, but with the additional assumption that emission
only occurs outside $r_{ms}$. The solid dots show $\chi^2$ for $\phi_c=\pi/2$
(case A1) while the open circles for A10.  The solid histogram shows the
resulting $a_s$ measured for 25 monte carlo simulations for A1; the dotted
line for A10.
\end{normalsize}
\vskip 3mm
\addtolength{\baselineskip}{3pt}
}

If the disk is much hotter than we have assumed, then the parameters
will not be as well constrained as those that we have used, as
we would then sample only the outer regions of the disk.  To
illustrate this point, we have run a somewhat unrealistic model, A13,
with the fiducial accretion rate, but a black hole mass of
$2\times10^8 M_\odot$ (near the Eddington limit).
A standard blackbody accretion disk around a black hole with this mass
cannot fit the observations as its spectrum is too steep and the
magnification must be much larger than in standard models of the lens
galaxy.  The error on the measured spin is much larger than for Run A1
(see Table 2).  The $U_i$ for Run A13 are comparable to those 
in Run A1, so the intensities are recovered similarly well.

To see how well we can perform the inversion when the assumption
of smooth radial variation is
incorrect, we multiplied the accretion disk intensity by
\begin{equation}\label{fluceq}
1+\sin{\left[6\pi {\log(r)-\log(r_{in})\over \log(r_{out})-\log(r_{in})}\right]},
\end{equation}
which makes the disk three logarithmically spaced annuli (runs A14 and A15).
Surprisingly, the recovered model parameters, $\bzeta$, are accurate.
Whether the radial variations can be discovered depends on the number of
observations.  In Run A14 (15 observations), the correct overall shape
is found, but the radial modulation not reproduced; in Run A15
(41 observations), the radial dependence of the recovered intensities
is more nearly correct.  This indicates that the inversion is only accurate
if our smoothness model assumption is met on the smallest scale probed
by the sampling.
Figure 15 shows the results of run A14.
\vskip 2mm
\hbox{~}
\centerline{\psfig{file=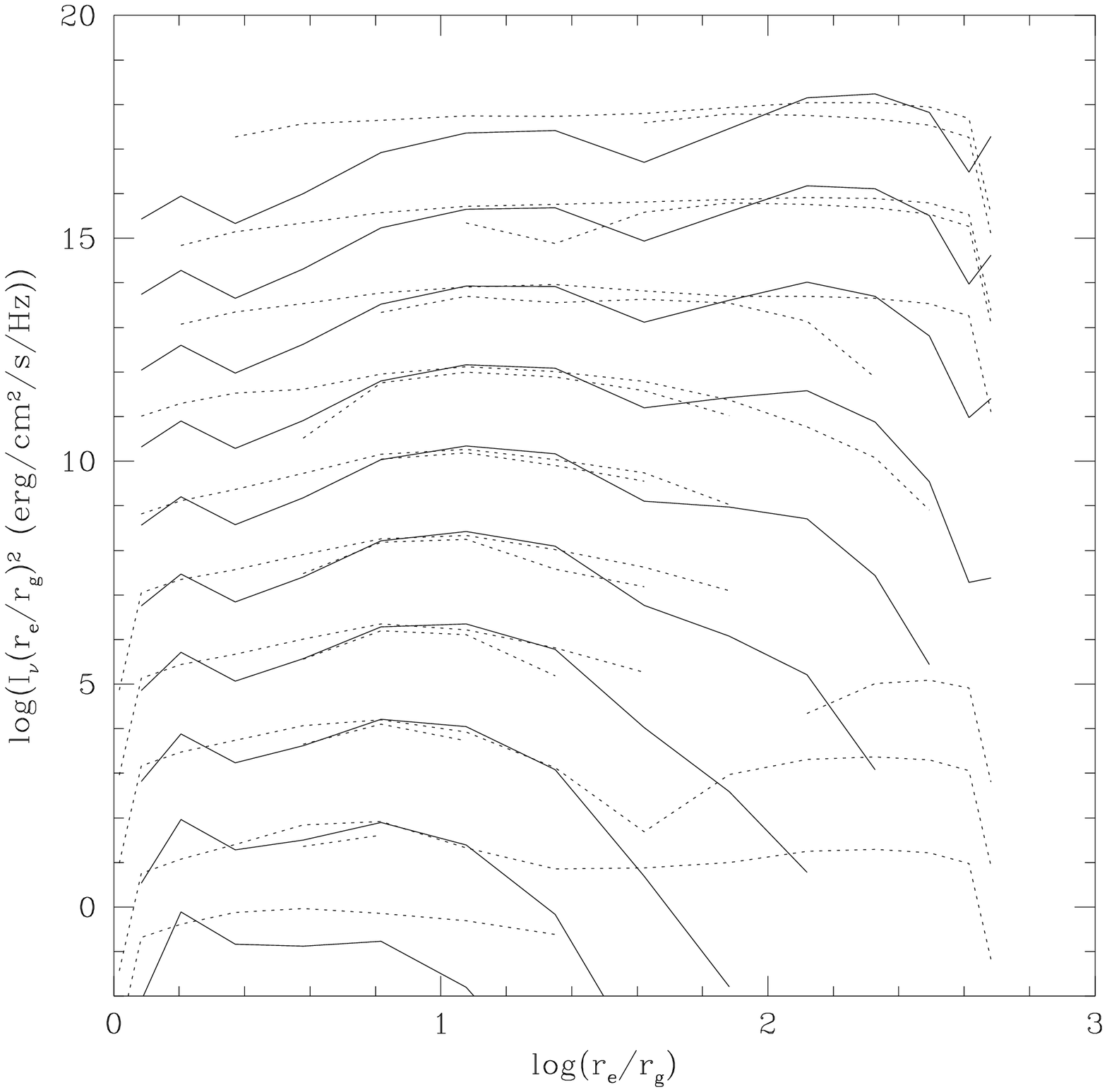,width=3.7in}}
\noindent{
\scriptsize \addtolength{\baselineskip}{-3pt}
\vskip 1mm
\begin{normalsize}
Fig.~15.\ Plot of original (solid lines) and recovered (dotted lines) for 
blackbody disk with fluctuations added (model A14).  Each curve shows a 
different frequency (from $3\times 10^{14}$ to $10^{16}$ Hz), shifted by 2 
units for clear separation.  The dotted lines show the maximum and minimum 
recovered intensity from 20 Monte Carlo realizations, with only positive 
intensities plotted.
\end{normalsize}
\vskip 3mm
\addtolength{\baselineskip}{3pt}
}
 
We have not tried breaking the assumption of azimuthal symmetry, as the
transfer function is computed assuming it.

\subsection{Varying observation parameters}

The question of how many observations are necessary is addressed with Runs
M2a-d (Table 2).  First, we compare fewer observations (21)
at the same sampling rate (M2a) and for the same duration (M2b).  In
each case, the rms scatter of the model
parameters is remarkably small compared to Run A1.  Thus, {\it if
the underlying model is correct}, it appears that we can determine
the model parameters with a high degree of accuracy with relatively
few observations.

A smaller number of observations, however, impairs our ability
to reconstruct the true surface brightness of the accretion disk,
as the inversion relies more upon the smoothing
constraint than on the actual data.  This can be seen in
Figure 16, which shows the derived formal accuracies for
various numbers of observations.
\vskip 2mm
\hbox{~}
\centerline{\psfig{file=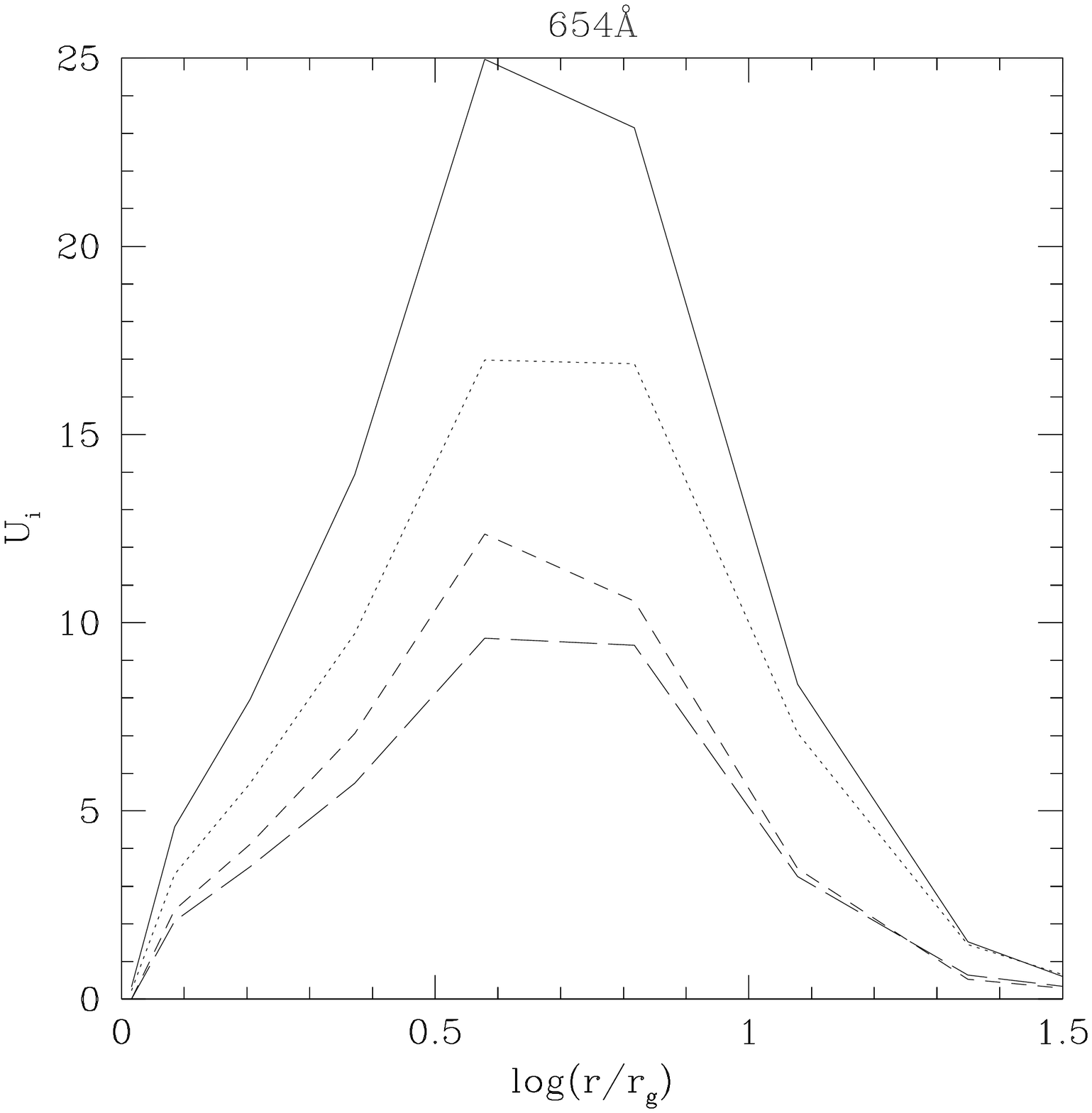,width=3.7in}}
\noindent{
\scriptsize \addtolength{\baselineskip}{-3pt}
\vskip 1mm
\begin{normalsize}
Fig.~16.\ This shows a plot of the $U_i$ as a function of radius for various 
numbers of observations at a single frequency: solid line is $N_t=41$ (A1), 
dotted line is $N_t=21$ (M2b), short-dashed line is $N_t=11$ (M2d), and 
long-dashed line is $N_t=5$ (M2e).
\end{normalsize}
\vskip 3mm
\addtolength{\baselineskip}{3pt}
}

   The total amount of information about the intensities that may be
gleaned also depends on the number of observations.  It is obvious
that the number of frequencies at which the intensity may be inferred
scales in proportion to the number of bands whose lightcurves are measured.
In addition, comparing A1 with M2b-d shows that the number of radial
points with reliable solutions increases slowly with increasing number of
observation times $N_t$.  This is because each observation constrains
most strongly the minimum radius where the caustic crosses. Since
the observations are spaced linearly, while the radii are spaced logarithmically,
the number of radial points constrained by the data is $\propto 
\log(r_{max}/r_{min}) \propto \log(T/\Delta t) \propto \log(N_t)$, where 
$T$ is the total duration and $r_{max}$ and $r_{min}$ are radii corresponding
to the radial limits of the region with a reliable solution.  In addition,
$U_i \propto N_t^{1/2}$ (see
Figure 16), in the usual fashion of signal-to-noise ratios.

In addition to the number of observations, the SNR of each observation
affects the quality of the inversion.  Runs M3a-b with SNR=50 demonstrate
this dependence.  For Run M3a, the recovered parameters have similar errors to 
those in Run A1, except for the errors on the spin.  For Run M3b, the errors on 
recovered parameters are roughly double those in Run M2d.  
For fixed model parameters and fixed number of observations, on average the
errors on ${\bf I}$ are directly proportional to the errors on ${\bf F}$;
in other words, $\langle U_i \rangle \propto SNR$, where the average is over 
20 simulations.  For individual simulations, the $U_i$ depend on $\lambda$, 
which causes scatter in $U_i$.

Next, we look at the dependence of the solution on which frequencies
are observed.  Run M4a shows that the best fit model parameters, 
$\bzeta$ are not strongly dependent on the infrared
frequencies.  However, excluding infrared bands reduces $U_i$ at the lowest 
frequencies.  Run M4b shows that the best fit $\bzeta$
are strongly dependent on observing the ultraviolet frequencies.
This run has the largest errors on model parameters and poorest agreement 
with the true
parameters.  In addition, the $U_i$ at the highest frequencies are strongly
reduced for this run. 
Comparing Runs A1, M4a,b, we find that the number of frequencies 
(at the accretion disk) constrained by the data $\propto N_\nu$, which shows
why broad frequency coverage is crucial for mapping out the disk spectrum.

Runs A1, A3, and A4 show that both the number of radial points with
reliable recovered intensities and $U_i$ are insensitive
to the total duration of the monitoring within the range of uncertainties
in $v_c$.  %Since we accept only $I_i > 0.1 I^{max}_i$,
%so that longer monitoring can not increase the sensitivity at larger radii
%where the intensities are smaller.
Runs A2 shows that the time at which we begin the observations does
not strongly affect the derived parameters or surface intensity, as
long as we catch the region near the peak.

\section{Discussion}

\subsection{Observational requirements}

    Experimental design depends on which questions are the goal, and to
what accuracy one aims to answer them.  Consequently, this discussion,
much like our earlier discussion of the experimental errors, divides
according to whether the primary aim is to map the surface brightness of
the disk, or to infer the model parameters.

   If the goal is to obtain a map of the the disk intensity as a
function of radius and frequency over as large region in the $r_e$--$\nu_e$
plane as is possible, it pays most to invest in
multiple monitoring bands because the number of radius-frequency pairs
for which it is possible to find a solution is $\propto N_\nu$, but
rises only logarithmically with $N_t$.  We expect the intensity to
vary as a power law in radius, so a better observation strategy
might be to space observations logarithmically in time; however, this
will be difficult to achieve in practice, which is why we have assumed
equal time spacing.  The accuracy of the solution is
directly proportional to the SNR in the data.
%since $N_U \bar U_i \propto N_t^{0.3} SNR N_\nu$.
To map the regions nearest to the black hole, the highest UV frequencies
are crucial (although this depends on the assumption that the spectral
peak moves to higher frequencies at smaller radii); however, lower
frequencies are required if one is interested in obtaining a broad-band
spectrum.  Note also that a higher sampling frequency
increases the chance that one will obtain monitoring data during 
the time when the caustic line is near the black hole.
To obtain high formal accuracy for a broad range in radius
requires a lightcurve that is finely sampled for a long time period.
To determine the intensities, the model parameters must also be well-defined.
When the observations are too few or the SNR too small, the uncertainties in
$\bzeta$ contribute to the uncertainty in $I_i$.

If the goal is to simply constrain the model parameters, and one believes that
the model is correct, then only a small number of observations may be required
(5 observations were sufficient in our simulations to constrain all model
parameters to better than 20\% at $1 \sigma$; 11 observations for 
$\leq$ 10\% $1\sigma$ accuracy).
Since the black hole mass and caustic velocity are quite uncertain, it may
be necessary to have more observations to be sure to obtain the few near the peak of the
caustic crossing that are essential for success.
The highest frequencies are most important for constraining model parameters.

In our simulations we assumed that only statistical errors affected the fluxes,
and that these errors can be estimated from the observational data.  There
might also be systematic errors in the fluxes due, for example, to emission
line contributions, inaccurate calibration between bands, or cosmic rays.
%Such unaccounted-for systematic errors may change the quality of the results
%presented here.
The quality of the result could also be affected by these problems.

\subsection{Under what circumstances might the model fail?}

   The inversion scheme we propose assumes a model that is plausible---a
geometrically thin relativistic accretion disk in which azimuthal
variations are quickly smoothed out, microlensed by a caustic system
whose basic length scale is much larger than the size of the bright
region of the disk.  However, we are by no means guaranteed that even
this general framework is correct.  In this section we discuss what would
happen if our method is attempted, but one of these assumptions is invalid.

   The most basic of our assumptions is that the surface brightness
varies smoothly as a function of radius.  Given sufficiently dense
sampling with good SNR, even quite sharp gradients could be recognized
by our procedure.  On the other hand, if the sampling is inadequate,
the existence of such features would appear only as a troubling inability
to find a solution with adequate $\chi^2$, unless $\lambda$ is taken to be
very small.

   Another potential source of trouble is departures from azimuthal
symmetry.  In the absence of microlensing, ``spots" can modulate
the lightcurve on the orbital period if they are in the relativistic
portion of the disk and the inclination is relatively large (Abramowicz et
al. 1991).  In this case, that would mean periods of $\simeq 31 M_9 
(r/10r_g)^{3/2}(1+z)/2.7$~days (in the observed frame).  Because the 
largest observed variations in the
Einstein Cross are $\sim 10\%$ on year-long timescales, this effect
cannot be too strong in this system.  However, there might be a range
of spot brightness in which they are too weak to show up in the
ordinary lightcurve, yet strong enough to cause some periodic modulation
of the lightcurve during a microlensing event (Gould \& Miralda-Escud\'e 1997).
Because the likely duration of a microlensing event (a few weeks) is
several to ten times the orbital period for the brightest part of the
disk, it is possible that this effect might be seen directly in the
lightcurve.  If not, they might still make it difficult to find a solution
with acceptable $\chi^2$.  Particularly if the break in azimuthal
symmetry is approximately $\propto e^{i\phi}$ and the disk is nearly
face-on, a strong spot might be confused with Doppler boosting, leading
to a mistaken inference for the disk inclination.

    Several physical effects might make disks geometrically thick---radiation
pressure support if the luminosity approaches Eddington (Abramowicz et al.
1988), gas pressure support if the ions retain most of their heat
(Rees et al. 1982, Narayan \& Yi 1995), or an optically thick outflow.
If any of these mechanisms acts, the orbital velocity at the photosphere
would no longer be that corresponding to circular free-fall in the
equatorial plane of the black hole, so that the general relativistic
transfer function we apply would no longer be valid.  We would expect,
then, difficulties in finding a solution with acceptable $\chi^2$, but
the portion of our solution describing the outer regions of the disk
should be only weakly affected.

   The surface brightness model we have used for the inversion simulations
is roughly consistent with both the observed spectrum (given the uncertainty in
reddenning) and the current microlensing size constraint, but it is not unique. 
If the actual spectrum has strong emission in the far ultraviolet (which could
be true if the reddening is greater than our estimate), the observable portion
of the spectrum will be dominated by emission far from the horizon, weakening
all the relativistic effects.  If so, the disk inclination and black hole
spin will be poorly constrained.

    Uncertain reddening can have other effects, also.  Because we can
expect it to be uniform across the face of the disk, it should not affect
the inferred radial profile of the disk, but it could well introduce
additional uncertainty into our estimate of the intrinsic disk spectrum
at any given radius.

    Our assumption that the bright part of the disk is small relative to the
caustic length scale is unlikely to be broken, except in the outer
regions of the accretion disk.  In this case, for which
the microlensing optical depth is $\sim 1$ (Witt \& Mao 1994), the caustic
scale is essentially the size of the Einstein ring due to a single star.
Consequently, the ratio between the disk size and the caustic scale
is only $\simeq 0.01 M_9 (r/10 r_g) m^{-1/2} (h/0.75)^{1/2}$, where $m$
is the mean mass of microlensing stars in Solar units, so according
to the Grieger et al. (1988) criterion, the caustic assumption is
likely to be valid out to 100 $r_g$.  In our fiducial model, for example,
the flux at $100 r_g$ peaks at an observed wavelength $1[(1+z)/2.7]\mu$m, 
where we have scaled to the redshift of the Einstein Cross.

      However, there might be difficulties in practice from a related
problem: measuring $A_0 F_\nu$. Again, if the disk size is much smaller
than the Einstein radius of a single star, then $A_0$ is approximately
constant during a high amplification event;  thus, the same criterion
for success applies as in the previous paragraph.  Of course, observations
when the quasar is outside the caustic are still required to measure
$A_0$.  Given the expectation that the quasar is smaller at higher
frequencies, observing at the highest frequencies will provide the best
constraint on $A_0$.  To determine how these difficulties and those listed 
in \S 1.2 will affect the inversion, we are currently running simulations 
of full microlensed lightcurves appropriate for the Einstein Cross, to
which we will apply our inversion algorithm (Wyithe \& Agol, in preparation).

     In any of these instances of model inappropriateness, the impact on
specific inferred parameters depends somewhat on details of the inversion
procedure.  For example, if the procedure we have outlined is followed
(i.e., minimizing $\chi^2$ by varying $\bzeta$ at fixed $\lambda$,
then raising $\lambda$ until $\chi^2$ meets our definition of acceptability),
difficulty in achieving satisfactory $\chi^2$ reduces the ultimate $\lambda$.
This means that, in effect, more degrees of freedom are ``spent" on fitting
the $I(r,\nu)$, leaving fewer for defining $\bzeta$.  If the problem
is lack of smoothness in the radial profile, this transfer of effort
is reasonable; if the problem is different, however, and if one cares
about the accuracy of the $\bzeta$ parameters, one might choose to
modify the procedure in a way that keeps $\lambda$ fixed at a relatively
large value.

      One disadvantage of using linear regularization is that the
intensities are not required to be positive definite, though it is 
impossible to emit a negative number of photons.
We have tried to incorporate this by trying three other methods:
maximum entropy; replacing $I_i$ by $log(I_i)$ in ${\cal B}$; and 
the method of projections onto convex sets. Each technique finds solutions
that are local minima with large $\chi^2$.  Another useful technique might 
be to make the further assumption that the spectrum at each radius can be 
described as a blackbody, and then solve for the 
temperature as a function of radius.  These methods are all non-linear, 
and thus intrinsically slow. They therefore impede the exploration of
parameter space, but might be interesting for future work.

\subsection{Multiple microlensing events}

Of the five parameters in $\bzeta$, three ($t_o$, $v_c$, and $\phi_c$)
will change from one event to the next, but the other two ($a_s$ and $i$)
should remain fixed.  Since we expect roughly one event per year, it
should be possible to combine observations of several events in order
to more tightly constrain $a_s$ and $i$.

\subsection{Connection to X-ray microlensing events}

   The optical and ultraviolet continua of quasars are not the only portions
of the spectrum radiated by the inner part of the accretion disk.  The X-ray
continuum must also come from somewhere near that region.  It, too, should
therefore be microlensed in much the same way as the optical and ultraviolet
continuum we have discussed in this paper.

    Whether the same technique can be successfully applied to the X-ray
continuum depends on the same considerations as discussed in \S 4.2, but
several of them are more likely to present problems in the context of
X-rays than for the optical/ultraviolet continuum.  There have been numerous
suggestions, for example, that the X-rays are produced in a relatively
small number of compact active regions (e.g., Haardt et al. 1994) that might
have substantial velocities relative to the disk (Beloborodov 1999).  If so,
the assumptions of azimuthal symmetry, radial smoothness, and also
simple circular orbital motion might all be suspect.

    Nonetheless, it would certainly be worthwhile to monitor the X-ray
flux during a microlensing event, in the hope that its emissivity distribution
is sufficiently consistent with our assumptions that it, too, could be
mapped.  Combining this data set with the optical/ultraviolet data would
also provide an independent constraint on the $\bzeta$ parameters, which
should all be the same for the same event.

\subsection{Summary}

     We have demonstrated that monitoring microlensing events in the
Einstein Cross quasar has great potential for both revealing the structure
of its continuum emission with unprecedented resolution, and potentially 
constraining such basic parameters of the quasar as the spin of its black 
hole (if we assume emission occurs only outside of the marginally stable
circular orbit), the mass of the black hole (modulo the 
caustic velocity), and the inclination angle of its disk relative to our 
line of sight.  If this potential is realized, and the analytic method we 
have proposed is implemented successfully, we may be able to begin answering 
such fundamental questions as:  What is the intrinsic local spectrum of the 
disk?  How close is it to thermal?  And, most fundamentally,
does the dissipation distribution in accretion disks vary with
radius in the fashion predicted (see equation~1) long-ago?

\acknowledgments

We would like to thank Andy Gould and Chris Kochanek for useful discussions.
We thank Casey Papovich, Keiichi Wada, Andrew Zirm, and Viktor Ziskin
for time on their workstations.  Thanks to David Heyrovsky and the
referee for comments which improved the manuscript.

This work was partially supported by NASA Grant NAG 5-3929 and NSF Grant
AST-9616922.

\pagebreak
\begin{deluxetable}{ccccccc}
\tablewidth{0pt}
\tablecaption{Model parameters for Monte Carlo runs (for $N_t=41$, SNR = 100, 
and $N_\nu$=11)}
\tablehead{\colhead{Model}
&\colhead{$t_0(\Delta t)$} & \colhead{$v_c(r_g/\Delta t)$} & 
\colhead{$\mu$} & \colhead{$\phi_c$} & \colhead{$a_s$} & \colhead{$\lambda$}
}
\startdata
A1 &  0    & 1   & 0.866   &$\pi/2$& 0.998    & \nl 
%   &    &     &   & & -.09 & .93 & 0.84    & 1.57   & {\it 1} & 1.1 \nl %run 115 [X]
&-.1$\pm$.2&.96$\pm$.06&.84$\pm$.02&1.56$\pm$.03&.96$\pm$.05&.01-1.9 \nl %run134stats [X]
\tableline
A2 & 5     & 1   & 0.866   & $\pi/2$ & 0.998  &    \nl 
%  & &    &     &   & & 4.7   & .99 & 0.83    & 1.6     & 0.95   & 0.8 \nl %run 116 [X]
&-4.8$\pm$.2&.96$\pm$.06&.84$\pm$.02&1.58$\pm$.02&.97$\pm$.03&.002-1.7 \nl %run154stats [X]
\tableline
A3 & 0     & 2   & 0.866   & $\pi/2$ & 0.998  &     \nl 
%   &    &     &   & & -.03  & 1.95& 0.83    & 1.56    & 0.91   & 0.9 \nl %run 113 [X]
&-.10$\pm$.13&1.95$\pm$.095&.83$\pm$.03&1.58$\pm$.03&.93$\pm$.08&.01-1.6 \nl %run152stats [X]
\tableline
A4 & 0     & .5  & 0.866   & $\pi/2$ & 0.998  &     \nl 
%   &    &     &   & & -.10  & .52 & 0.86    & 1.58    & 0.98   & 1.1 \nl %run 131 [X]
&-.18$\pm$.2&.50$\pm$.02&.85$\pm$.02&1.57$\pm$.02&.97$\pm$.03&.001-2.3 \nl %run157stats [X]
\tableline
A5 & 0     & 1   & 1.0     &  n/a    & 0.998  &     \nl 
%   &    &     &   & & .07   & 1.14&{\it 1.0}&{\it 0}  & {\it 0}& 1.2 \nl %run 114 [X]
&-.06$\pm$.07&1.02$\pm$.04&.999$\pm$.001&1.53$\pm$.08&.99$\pm$.02&.0001-1.8\nl %run153stats [X]
\tableline
%the following line is a 60 degree viewing angle
A6 & 0     & 1   & 0.5   & $\pi/2$ & 0.998    &     \nl
%   &    &     &   & & .08   & .93 & .51   & 1.55    & {\it 1}  & 1.0 \nl %run 120 [X]
&-.03$\pm$.08&.95$\pm$.05&.50$\pm$.02&1.57$\pm$.01&.92$\pm$.06&.005-1.8 \nl %run155stats[X]
\tableline
A7  & 0     & 1   & 0.866   & $\pi/4$ & 0.998  &     \nl
%   &    &     &   & & -.06  & 1.0 & 0.84    &  .83    & 0.88   & 1.0 \nl %run 132 [X]
&-.02$\pm$.13&.99$\pm$.04&.84$\pm$.01&.83$\pm$.05&.95$\pm$.11&.01-1.7 \nl %run156stats [X]
\tableline
A8  & 0     & 1   & 0.866   & $3\pi/4$& 0.998  &     \nl
%   &    &     &   & &  .02  & 1.0 & 0.85    & 2.33    & {\it 1}& 1.1 \nl %run 133 [X]
&-.03$\pm$.1&1.01$\pm$.03&.86$\pm$.01&2.33$\pm$.04&.99$\pm$.02&.003-1.9  \nl %run158stats [X]
\tableline
A9 & 0     & 1   & 0.866   & $3\pi/2$ & 0.998    &     \nl
%   &    &     &   & & .08   & .93 & .51   & 1.55    & {\it 1}  & 1.0 \nl %run 160 
&.1$\pm$.1&.98$\pm$.06&.84$\pm$.02&4.72$\pm$.02&.95$\pm$.05&.006-1.9 \nl %run160stats[X]
\tableline
A10  & 0     & 1   & 0.866   & $\pi$ & 0.998    &     \nl
%   &    &     &   & & -.27  & 1.1 & 0.84    & 3.37  & 0.97     & 1.0 \nl %run 123 [X]
&.03$\pm$.07&1.03$\pm$.03&.86$\pm$.01&3.16$\pm$.08&.99$\pm$.03&.005-1.8 \nl %run162stats [X]
\tableline
A11a  & 0     & 1   & 0.866 & $\pi/2$ & 0.       &     \nl 
%   &    &     &   & & -.40  & .85 & 0.81  & 1.58    & .55      & 0.4 \nl %run 111 [X](no r constraint)
&-.46$\pm$.26&.95$\pm$.07&.78$\pm$.06&1.57$\pm$.02&.17$\pm$.32&.005-.8  \nl %run149stats[X]
%   &    &     &   & & -.02  & .78 & 0.91  & 1.57    & .41      & 1.9  \nl %run 112 [X](constraint r>r_ms)
A11b\tablenotemark{a}&-.22$\pm$.24&.87$\pm$.07&.87$\pm$.02&1.57$\pm$.02&.09$\pm$.12&.001-4.4 \nl %run150stats[X] NOTE:  The correlation is not nearly so good in this case - histograms show large
%systematic errors.  Also, N_U can't be compared since radius grid is different.
% to show what happens when we constrain r> rms
\tableline
A12 & 0     & 1   & 0.866   & $\pi/2$ & 0.5    &     \nl
%   &    &     &   & & .08   & .93 & .51   & 1.55    & {\it 1}  & 1.0 \nl %run 159 [X]
&-.29$\pm$.20&.91$\pm$.09&.83$\pm$.07&1.57$\pm$.02&.55$\pm$.25&.03-.7 \nl %run159stats[X]
\tableline
\tablenotetext{a}{We have constrained $r > r_{ms}$.}
\enddata \label{tab1}
\end{deluxetable}
\begin{deluxetable}{ccccccccccc}
%\tablewidth{6.5in}
\tablewidth{0pt}
\tablecaption{Monte Carlo parameters (for $t_0=0$, $\mu =.866$, $\phi_c=\pi/2$, $a_s=.998$ 
model)}
\tablehead{\colhead{\#} & \colhead{$N_t$}& \colhead{SNR} &
\colhead{$N_\nu$} & \colhead{$v_c$} & \colhead{$v_c$(meas)}
&\colhead{$t_0(\Delta t)$} &
\colhead{$\mu$} & \colhead{$\phi_c$} & \colhead{$a_s$} & \colhead{$\lambda$}
}
\startdata
%A13\tablenotemark{a} & 11 & 100 & 11 & 4 &3.3$\pm$.5&-.1$\pm$.2&.79$\pm$.08&1.56$\pm$.07&.88$\pm$.15&1.3-15 & \nl %run147stats [X]
A13\tablenotemark{a} & 41 & 100 & 11 & 1 &.93$\pm$.07&-.2$\pm$.4&.85$\pm$.04&1.58$\pm$.06&.89$\pm$.19&.01-2 \nl %run168stats [X]
\tableline
A14\tablenotemark{b} & 15 & 100 & 11 & 2.9 &2.6$\pm$.3&-.04$\pm$.06&.90$\pm$.04&1.57$\pm$.03&.87$\pm$.19&.07-.9 \nl %run145stats [X]
\tableline
A15\tablenotemark{b} & 41 & 100 & 11 & 1 &.98$\pm$.06&.05$\pm$.20&.88$\pm$.05&1.57$\pm$.02&.79$\pm$.20&.002-.18 \nl %run151stats [X]
%\tableline
%A16\tablenotemark{c} & 41 & 100 & 11 & 1 &.96$\pm$.05&-.15$\pm$.15&.83$\pm$.02&1.61$\pm$.02&.97$\pm$.03&.002-1.6 \nl %run165stats [X]
\tableline
M2a & 21 & 100 & 11 & 1 &.99$\pm$.04& -.13$\pm$.15&.84$\pm$.01&1.58$\pm$.02&.98$\pm$.05&.5-5.2\nl %run135stats[X]
%   &    &     &   & & -.13 & .95 & 0.85  & 1.58    & {\it 1}  & 3.7 \nl %run 119 [X]
M2b & 21 & 100 & 11 &  2  &1.9$\pm$.13&-.12$\pm$.07&.82$\pm$.02&1.59$\pm$.03&.93$\pm$.07&.2-4.8  \nl %run138stats[X]
M2c & 15 & 100 & 11 &  2.9 &2.7$\pm$.3&-.09$\pm$.1&.83$\pm$.03&1.59$\pm$.03&.95$\pm$.06&.6-6.6 \nl %run139stats [X]
M2d & 11 & 100 & 11 & 4 &3.9$\pm$.3&-.03$\pm$.07&.83$\pm$.03&1.56$\pm$.03&.9$\pm$.1&1.7-11 \nl %run140stats [X]
M2e & 5 & 100 & 11 & 10 &$8\pm2$&0.$\pm$.15&.75$\pm$.1&1.48$\pm$.28&.93$\pm$.11&9-38 \nl %run169stats [X]
\tableline
M3a & 41 & 50  & 11 & 1  &.94$\pm$.07&-.25$\pm$.22&.83$\pm$.03&1.58$\pm$.03&.94$\pm$.14&.006-1.1\nl %run163stats [X]
%   &    &     &   & & -.18  & 1.03& 0.89  & 1.58  & {\it 1}    & .07 \nl %run 117 [X]
M3b & 11 & 50  & 11 &4 &3.9$\pm$.3&-.03$\pm$.14&.81$\pm$.06&1.55$\pm$.1&.88$\pm$.17&.8-8.6 \nl %run148stats [X]
%M3c & 41 & 400  & 11 & 1 &1.00$\pm$.02&-.01$\pm$.04&.86$\pm$.01&1.571$\pm$.006&.996$\pm$.003&3.e-5-1.8 \nl %run161stats [X]
\tableline
M4a & 15 & 100 & 8 & 2.9 &2.8$\pm$.24&-.04$\pm$.08&.84$\pm$.03&1.57$\pm$.03&.9$\pm$.1&.004-5.\nl %run146stats[X]
M4b & 41 & 100 & 4 & 1&.88$\pm$.09&-.6$\pm$.4&.80$\pm$.06&1.61$\pm$.05&.85$\pm$.14&.25-3.3  \nl %run144stats [X]
%   &    &     &   & & -.71 & .85 & 0.78  & 1.59  & 0.967      & 2.5 \nl %run 118 [X]
\tableline
\tablenotetext{a}{This run has $M_9=0.2$.}
\tablenotetext{b}{The intensities have been multiplied by equation (\ref{fluceq}).}
\enddata \label{tab2}
\end{deluxetable}

\begin{references}
Abramowicz, M., Bao, G., Lanza, A.  \& Zhang, X. 1991, A \& A 245, 454

Abramowicz, M., Czerny, B.,
Lasota, J.-P. \& Szuszkiewicz, E. 1988, \apj\ 332, 646

Agol, E., 1997, PhD Thesis, University of California, Santa Barbara
 
Albrow, M. et al., the PLANET Collaboration, 1998, ApJ submitted, astro-ph/98114
79
 
Bardeen, J. M. \& Petterson, J. A., 1975, ApJ, 195, L65
 
Beloborodov, A. M., 1998, MNRAS, 297, 739
Beloborodov, A. 1999, astro-ph/9809383
 
Corrigan, R.T. et al., 1991, AJ, 102, 304
 
Czerny, B., Jaroszy\'nski, M., \& Czerny, M., 1994, MNRAS, 268, 135
 
Cunningham, C. T., 1975, ApJ, 202, 788
 
Gaudi, B. S. \& Gould, A., 1998, ApJ, submitted, astro-ph/9802205
 
Gould, A. \& Miralda-Escude, J., 1997, ApJ, 483, L13
 
Gould, A. \& Gaudi, B. S., 1997, ApJ, 486, 692

Grieger, B., Kayser, R., \& Schramm, T., 1991, A \& A, 252, 508
 
Grieger, B., Kayser, R., \& Refsdal, S., 1988, A \& A, 194, 54
 
Haardt, F., Maraschi, L. \& Ghisellini, G.  1994, \apjl\ 432, L95

Huchra, J., et al., 1985, AJ, 90, 691a
 
Irwin, M. J., et al., 1989, AJ, 98, 1989
 
Jaroszy\'nski, M. \& Marck, J.-A., 1994, A \& A, 291, 731
 
Krolik, J. H., 1998, {Active Galactic Nuclei}, (Princeton: Princeton University
Press)

Jaroszy\'nski, M., Wambsganss, J., Paczy\'nski, B., 1992, ApJ, 396, L65

Narayan, R. \& Yi, I., 1994, ApJ, 428, L13

Page, D. N. \& Thorne, K. S., 1974, ApJ, 191, 499

Press, W., Teukolsky S., Vetterling W., \& Flannery, B., 1992, 
{\it Numerical Recipes in FORTRAN}, 2nd edition, (Cambridge: Cambridge
University Press)

Pringle, J. E., 1981, ARAA, 19, 137

Rauch, K. P., \& Blandford, R. D., 1991, ApJ, 395, L65

Rauch, K. P., \& Blandford, R. D., 1994, ApJ, 421, 46

Rees, M. J., Phinney, E. S., Begelman, M. C., \&
Blandford, R. D., 1982, Nature, 295, 17

Schneider, P., Ehlers, J., \& Falco, E., 1992,
{\it Gravitational Lenses} (New York: Springer-Verlag)

Schneider, P. \& Wei\ss, A., 1986, A \& A, 164, 237

Wambsganss, J., Paczy\'nski, B., \& Schneider, P., 1990, ApJ, 358, L33

Wisotzki, L., K\"ohler, T., Kayser, R., and Reimers, D., 1993, A \& A, 278, L15

Witt, H. J., Kayser, R., \& Refsdal, S., 1993, A \& A, 268, 501

Witt, H. J. \& Mao, S., 1994, ApJ, 429, 66

Witt, H. J., Mao, S., \& Schechter, P. L., 1995, ApJ, 443, 18

Wyithe, J. S. B., Webster, R. L., \& Turner, E. L., 1999, astro-ph/9901341

\end{references}
\end{document}